\documentclass[prb,twocolumn,superscriptaddress,10pt,floatfix]{revtex4-2}
\usepackage{amsmath}
\usepackage{amssymb}
\usepackage{graphicx}
\usepackage{bm}
\usepackage{mathrsfs}
\usepackage{xcolor}
\usepackage[normalem]{ulem}
\usepackage{siunitx}
\usepackage{nicematrix}
\usepackage[colorlinks=true,allcolors=blue]{hyperref}

\def\be{\begin{equation}}
\def\ee{\end{equation}}
\def\bea{\begin{eqnarray}}
\def\eea{\end{eqnarray}}
\def\bs{\begin{split}}
\def\es{\end{split}}
\def\ni{\noindent}

\def\bi{\begin{itemize}}
\def\ei{\end{itemize}}
\def\f{\frac}

\def\mbf{\mathbf}

\def\a{\alpha}

\def\vare{\varepsilon}
\def\s{\sigma}
\def\d{\delta}

\def\o{\omega}

\makeatletter
\renewcommand\paragraph{\@startsection{paragraph}{4}{\z@}
  {3.25ex plus 1ex minus .2ex}{1.5ex plus .2ex}{\normalfont\normalsize\itshape\centering}}
\makeatother

\begin{document}

\title{Beyond Allen-Heine-Cardona: non-perturbative electron-phonon interactions in the linewidths and lineshifts of diamond}
\author{Jean Paul Nery}
\altaffiliation{nery.jeanpaul@gmail.com}
\affiliation{Nanomat group, QMAT research unit, and European Theoretical
Spectroscopy Facility,  Université de Liège, B5a allée du 6 août, 19,
B-4000 Liège, Belgium}
\author{Samuel Longo}
\affiliation{Nanomat group, QMAT research unit, and European Theoretical
Spectroscopy Facility,  Université de Liège, B5a allée du 6 août, 19,
B-4000 Liège, Belgium}
\author{Matthieu J. Verstraete}
\altaffiliation{matthieu.verstraete@uliege.be}
\affiliation{Nanomat group, QMAT research unit, and European Theoretical
Spectroscopy Facility,  Université de Liège, B5a allée du 6 août, 19,
B-4000 Liège, Belgium}
\affiliation{ITP, Physics Department, Utrecht University, 3508 TA Utrecht, The Netherlands}

\begin{abstract}
The temperature-dependent band gap of solids is usually computed from the
perturbative Allen-Heine-Cardona (AHC) electron-phonon self-energy
evaluated on-shell. Extending AHC to arbitrary frequency $\o$ to determine the full spectral function via the
Dyson equation is known to fail, misplacing satellites and yielding no
broadening at band extrema, while non-perturbative supercell (SC) methods
have focused on eigenvalue averages rather than lineshapes, and
approaches based on special displacements cannot describe the full lineshape,
or the lineshift at degenerate bands.
Here we use a non-perturbative Green's function method (NPG), stochastically sampling distorted SC configurations, from which the spectral function, including lineshift, linewidth, and asymmetry, follows directly,
and we recover finite spectral weight at the renormalized band extrema. We
prove that the perturbative self-energy, computed to any order with the
bare propagator and introduced into the Dyson equation, has an imaginary
part that vanishes within the bare gap, giving incorrect spectral
functions: self-consistency of the propagator is essential to broaden the band edges. NPG satisfies this property by construction, and contains all
non-bubble diagrams. We also give a simple explanation of why SC methods
converge with much smaller SCs than the corresponding $\mbf{q}$-grids required by perturbation theory.
For the band gap shift itself, the NPG and on-shell AHC results are found
to be comparable, demonstrating that higher-order terms do not
significantly alter the resulting renormalization in diamond.
When it comes to the spectral function though, our results show that going
beyond bare perturbation theory is not merely more accurate, but
necessary, and NPG provides a robust framework to capture spectral
broadening and higher-order effects from first principles.
\end{abstract}

\maketitle

\section{Introduction}

Electron-phonon interactions (EPIs) play a central role in condensed matter physics~\cite{Giustino2017}. They renormalize the band gap at 0~K due to zero-point motion and determine its temperature dependence, an effect that can range from a few meV up to hundreds of meV in materials like diamond, and can even reach 2 or 3 eV in molecular phases of hydrogen and other systems~\cite{Melo2023,Monacelli2021}. EPIs also determine the intrinsic linewidth of electronic states, which controls carrier mobilities and spectral broadening.
The full and quantitative inclusion of EPIs is therefore of central importance, not only for accuracy in the design of novel materials, but also from a fundamental point of view~\cite{Nery2022}, to determine the relative role of electron-electron interactions (correlations), EPIs, thermal expansion, anharmonicities, and other effects~\cite{Girotto2023,Girotto2025}.

Historically, computations of the electron-phonon contribution lacked a cohesive theoretical framework. Instead, researchers tried to focus on specific physical mechanisms, evaluating either how thermal atomic displacements altered the periodic lattice potential, or how electrons dynamically scattered off lattice vibrations. Allen, Heine, and Cardona (AHC)~\cite{Allen1976,Allen1981,Allen1983} unified the existing approaches and placed the theory on a firm foundation using diagrammatic perturbation theory (PT). In this framework, the Fan and Debye-Waller (DW) terms, originally used in isolation, are summed to provide the standard self-energy contributions used to compute the lineshift and linewidth.
The AHC method has been very successful as a clear and systematic framework to describe the temperature dependence of electronic bands in semiconductors, as well as the lifetime of electrons due to interaction with phonons. In addition, the underlying diagrammatic perturbation theory established in these early works can, in principle, be systematically expanded to include higher-order Feynman diagrams~\cite{Allen1978, Allen1981}.

The AHC approach only includes the lowest order diagrams, assuming that the interactions are weak and the atomic displacements are harmonic. If the electron-phonon coupling is strong, higher-order diagrams constructed from the standard first and second derivatives of the potential become necessary. Furthermore, when the harmonic approximation fails, for example for large displacements,
the theory must also incorporate intrinsic lattice anharmonicity and higher-order variations of the electron-ion potential~\cite{Lee2020,Wu2020}.
In the last decade, many of these shortcomings of AHC perturbation theory have become apparent. 
In particular, clear deficiencies 
emerge when computing spectral functions. In AHC, the self-energy is determined from the bare $G_0$ electronic line, as opposed to the full $G$.
So the self-energy is fixed, and there is a unique way to obtain the spectral function through the Dyson equation. We will refer to this approach as Dyson-AHC, also referred to as Dyson-Migdal in the literature.
In polar semiconductors, where strong EPIs produce secondary structures (satellites) alongside the main quasiparticle (QP) peak, Dyson-AHC predicts an incorrect distance between the QP peak and the satellite. This inadequacy is not restricted to polar materials: non-polar semiconductors also exhibit weaker ``nonpolaron" satellite structures~\cite{Abreu2022} which are also incorrectly placed. In addition, Dyson-AHC produces a poor lineshift~\cite{Nery2018}. As we demonstrate below, both issues are related to an inherent structural deficiency: the lack of self-consistency of this one-shot approach.

New approaches have emerged attempting to improve AHC. These include the cumulant expansion~\cite{Nery2018,Abreu2022,Verdi2017,Zhou2019}, diagrammatic quantum Monte Carlo~\cite{Mishchenko2000}, self-consistency~\cite{Lihm2024,Lihmetal2024}, and supercell (SC) non-perturbative approaches~\cite{Zacharias2020}, sometimes combined with phonon methods that include anharmonicities~\cite{Zacharias2020b,Zhou2019}.
The cumulant expansion~\cite{Nery2018,Abreu2022,Verdi2017,Zhou2019} incorporates higher-order diagrams in an approximate manner. It extends the range of application to intermediate couplings, but is insufficient in the very strong coupling regime. Works introducing the self-consistent Born approximation, which resums higher-order rainbow (Fan-like) diagrams, have started
to appear~\cite{Lihm2024,Lihmetal2024}. 
While promising, these implementations remain computationally demanding and the Born approximation does not include vertex corrections. In lightly doped semiconductors, Migdal's approximation breaks down and vertex corrections are important \cite{Mishchenko2014}. To address this issue, other first-principles works have directly evaluated fourth-order diagrams~\cite{Lee2020}, but they are approximately $10^4$--$10^5$ times more costly than the second-order Fan and DW terms. Finally, diagrammatic quantum Monte Carlo includes all diagrams stochastically, though the computational cost is exceptionally high. While historically limited to models like the Fröhlich Hamiltonian~\cite{Mishchenko2000}, it has very recently been extended to first-principles self-energies by using a data-compression technique that reduces memory and computational costs~\cite{Luo2025}.

Another limitation of AHC
is that, by construction, it cannot broaden the band extrema at $T=0$: the renormalized quasiparticle peak, which lies inside the bare gap, remains a sharp delta function. This is known at lowest order~\cite{Lihm2024}, but is actually a pathology of PT to any order, which we will prove in Sec.~\ref{sec:LimsAHC}. 
Thus, methods that rely only on PT cannot produce an imaginary part inside the gap.
A final practical limitation of AHC comes from a standard approximation to the DW term. Because exact second-order derivatives of the electron-ion potential are usually unavailable in current implementations of density functional perturbation theory (DFPT), computing the DW contribution requires imposing the rigid ion approximation and combining it with the acoustic sum rule, to express the second-order terms purely via first-order derivatives. This has been reported to introduce an error of about 5--10\%~\cite{Ponce2025}, but statistics on different materials are limited.

An alternative paradigm is to consider SC and finite thermal displacements of
atoms. In this case band renormalizations involve all orders of EPIs and
anharmonic couplings are included automatically, making the method non-perturbative. Such calculations naturally bypass the technical bottleneck of evaluating
the DW terms, and sidestep the question of self-consistency altogether,
since they build $G$ directly rather than through $\Sigma$. In our case we
do define a self-energy, but \emph{from} $G$ rather than the other way
round, so it is a functional of the full propagator by construction. Such calculations
can therefore produce the spectral weight at the band edges that
Dyson-AHC cannot.

Non-perturbative SC-based calculations are computationally expensive, as DFT
scales $O(N_{el}^3)$ with system size. To circumvent this, the special
displacement method~\cite{Zacharias2020} has become popular because it relies
on a single optimized SC calculation. It is based on setting the phonon
amplitudes to their root-mean-square values ($\sigma_{\mathbf{q}s}$) and
assigning alternating signs to nearby momenta. This phase alternation attempts
to maximize error cancellation, approximating the results of a stochastic
ensemble average using only a single configuration.
In the infinite SC limit many properties are exact, but realistic SC sizes are quite limited in practice for DFT ($5^3$ to $10^3$ indicatively).
Relying on only one configuration has the following limitations: (i) It does not 
account for symmetries in calculating the splitting of degenerate defect levels~\cite{Kundu2024}; (ii) It fails to fully capture spectral broadening;
(iii) Evaluating a single deterministic displacement misses the statistical Wick contractions required for the diagonal pieces of higher-order diagrams. This yields incorrect combinatorial coefficients (e.g.\, returning $\sigma^4$ rather than $3\sigma^4$ in the harmonic approximation), though this discrepancy is restricted to a small region of phase space; (iv) Furthermore, evaluating observables at a single displacement intrinsically assumes that the potential $V$ varies strictly quadratically with ionic displacements, meaning higher-order derivatives of the potential cannot be captured. Although this can be mitigated using effective harmonic phonons that incorporate anharmonicities~\cite{Errea2013,TDEP2024} for the distortions, the formal issue remains. (v) Finally, generating the configuration requires mapping a one-dimensional sequence of alternating signs onto a three-dimensional Brillouin zone grid.
Certain adjacent momenta will inevitably share the same sign, preventing the complete cancellation of off-diagonal cross-terms.
Although (iii), (iv), and (v) might only introduce minor quantitative discrepancies, (i) and (ii) represent more serious drawbacks.

Previously, one of us developed a non-perturbative Green's function method~\cite{Nery2022}, which we will refer to as NPG, based on averaging over stochastically generated SC configurations. It was shown there that, in the adiabatic limit, this method exactly accounts for infinitely many Feynman diagrams. Furthermore, by formally defining a self-energy, we improved the accuracy of the standard expression used to calculate the spectral function, and illustrated the effect in a tight-binding model~\cite{Nery2022}. Although the Monte Carlo sampling in NPG in principle requires evaluating many configurations, particularly to resolve spectral widths, leveraging modern GPU clusters accelerates \textit{ab-initio} ground-state calculations by a factor of ten or twenty~\cite{GPU}, overcoming the prohibitive computational bottlenecks associated to CPUs and making DFT NPG calculations tractable.

In this work, we implement NPG within a first-principles framework using the ABINIT code~\cite{verstraete25} and apply it to diamond. This material is an ideal candidate to apply NPG: 
it exhibits a well-known large zero-point renormalization, so
higher-order PT effects are expected to be relevant; and it is widely
recognized as a highly harmonic system, so we can safely neglect lattice
anharmonicities.
Since diamond has an indirect gap, we distinguish throughout between the conduction band minimum (CBM), and the lowest conduction state at $\Gamma$, which we label
CB$_\Gamma$. 
In Sec.~\ref{sec:theory}, we begin by reviewing the theoretical formalism of NPG, and briefly outline the essential AHC equations. We describe the limitations of AHC compared to NPG, in particular when it comes to describing the imaginary part of the Green's functions, we prove that PT fails to give the self-energy an imaginary part inside the bare gap, and illustrate it with a simple model. In addition, we describe the conditions under which off-diagonal degenerate bands may have important non-perturbative effects.
Next, in Sec.~\ref{sec:convergence}, since we are doing the first implementation of NPG from first-principles, we do a detailed convergence study. Then in Sec.~\ref{sec:results}, 
we present converged first-principles NPG spectral functions and contrast our results against standard AHC theory at $T=0$. In addition, we compare the temperature dependence of the fundamental band gap renormalization between NPG and AHC, and explain why NPG does not need as dense grids as those of AHC. 
Finally, in Sec.~\ref{sec:Conclusions}, we summarize our findings.

\section{Theory}
\label{sec:theory}

\subsection{Non-perturbative method}

Here we introduce the basics of NPG~\cite{Nery2022}. This approach consists of the following steps: 1) stochastically generating distorted SC ionic configurations with the quantum phonon distribution, 2) calculating the electronic energies and wavefunctions of the ground state of each configuration, 3) calculating a Green's function for each of them, and then 4) averaging. Using harmonic phonons, the probability distribution is

\be
\scalebox{0.85}{$\displaystyle
P(\{u_{li\a}\}) = A \exp\left(-\sum_{\substack{li, mj \\ \alpha\beta, \mbf{q}s}}
\frac{\sqrt{M_i M_j}\,\o_{\mbf{q}s}}{2n_{\mbf{q}s} + 1}
\mathcal{E}^{li\alpha}_{\mbf{q}s} \mathcal{E}^{mj\beta}_{\mbf{q}s}
u_{li\alpha} u_{mj\beta}\right).$}
\label{eq:proba}
\ee

\ni Ionic configurations follow displacements $\{u_{li\a}\}$, with $l$ the index of the primitive cell (PC) within the SC, $i$ is the atom index in the PC, and $\alpha$ the Cartesian index.
The phonon momenta $\mbf{q}$ are commensurate with the SC, $s$ are the phonon branches, $\o_{\mbf{q}s}$ are the phonon frequencies, $n_{\mbf{q}s}$ the Bose-Einstein occupation factor, $M_i$ is the mass of atom $i$ (in this work just the mass of the carbon atom), $\mathcal{E}^{li\a}_{\mbf{q}s}$ are the (real) SC polarization vectors and can be written in terms of the polarization vectors of the PC, and $A$ is just the normalization factor.

Let $\{|\mathbf{k}n\rangle\}$ be a set of electronic eigenstates of the Hamiltonian of the undistorted cell with eigenvalues $\{\vare_{\mbf{k}n}\}$, $H_0 |\mbf{k} n \rangle = \vare_{\mbf{k}n} |\mbf{k}n\rangle$.
Let also $\{|J\rangle^I\}$ and $\{\vare_J^I\}$ be a set of eigenstates and eigenvalues at $\Gamma$ of configuration $I$, $H^I |J\rangle^I = \vare^I_J |J\rangle^I$. For additional technical details see Refs.~\cite{Allen2013,Nery2022}. The Green's function operator for each distorted configuration is defined as
 
\be
G^I = \f{1}{\o + i \d - H^I},
\label{eq:GI}
\ee 
 
\ni where $H^I$ is the distorted Hamiltonian. We are interested in the matrix elements in the basis of the undistorted eigenstates, $\langle \mbf{k}n| G^I(\o) |\mbf{k}n' \rangle$. Since we will compute the SC ground state at $\Gamma$, this provides access to all $\mbf{k}$ commensurate with the SC. Inserting the identity $\sum_J |J\rangle \langle J|$ results in~\cite{Allen2013,Nery2022}

\be
G^I_{\mathbf{k},nn'} = \sum_J \f{\langle \mathbf{k}n | J \rangle^I {}^I\langle J | \mathbf{k} n' \rangle}{\o + i \d - \vare^I_J}.
\label{eq:Ginit}
\ee

\ni The final Green's function is simply the average of the configurations,

\be
\tilde{G}_{\mathbf{k},nn'} = \lim\limits_{N_\mathrm{c} \rightarrow \infty} \f{1}{N_\mathrm{c}} \sum_{I=1}^{N_\mathrm{c}} G^I_{\mathbf{k},nn'}.
\label{eq:average}
\ee

Omitting the matrix band indices, one can define a self-energy $\Sigma_\mbf{k}$ through the Dyson equation

\be
\tilde{G}_\mbf{k}(\o + i \d) =  
\f{1}{G_{0,\mbf{k}}^{-1}(\o + i \d) - \Sigma_\mbf{k}(\o + i\d) },
\label{eq:Sigma}
\ee

\ni where $G_{0,\mbf{k}nn'}^{-1}(\o + i \d) = (\o + i \d - \vare_{\mbf{k}n})\d_{nn'}$ is diagonal. 
This self-energy contains an exact infinite-order summation of irreducible Feynman diagrams, incorporating all multi-phonon emission and absorption events along the electron propagator. If interactions are strong, the spectral function (obtained from the imaginary part of the Green's function) calculated in this way should have noticeable differences relative to AHC.

Although in principle the number of configurations is infinite in Eq.~\eqref{eq:average}, in practice it must be calculated with a finite set. Similarly, $\delta$
is formally a positive infinitesimal, but numerically this would require infinitely large SCs. 
A common strategy is to keep $\d$ at finite values and to converge with respect to it. A problem with this strategy, however, is that it also includes $\d$ in the unperturbed part of the Green's function, namely $G_0$ in Eq.~\eqref{eq:Sigma}, while there should be no broadening in the non-interacting term. To avoid this issue, 
we define a Green's function with $\d$ only in the self-energy:

\be
\begin{split}
G_\mbf{k}(\o) & = \f{1}{G_0^{-1}(\o) - \Sigma_\mbf{k}(\o + i\d) } \\ 
& = \f{1}{\o - \vare_\mbf{k} - \Sigma_\mbf{k}(\o + i\d)}.
\end{split}
\label{eq:final}
\ee

\ni This requires the self-energy to have a finite imaginary part, which
is generally the case in solids. We use this Green's function throughout
this work when applying NPG.
The spectral function is just given by the trace of the imaginary part of $G_\mbf{k}$,

\be
A_\mbf{k}(\o) = -\f{1}{\pi} \sum_n \Im \mathrm{m} G_{\mbf{k},nn}.
\label{eq:A}
\ee

\ni Our numerical procedure involves then computing Eqs.~\eqref{eq:Ginit} to \eqref{eq:A}.

\subsection{Plane waves}

\begin{figure}
    \centering
    \includegraphics[width=0.8\linewidth]{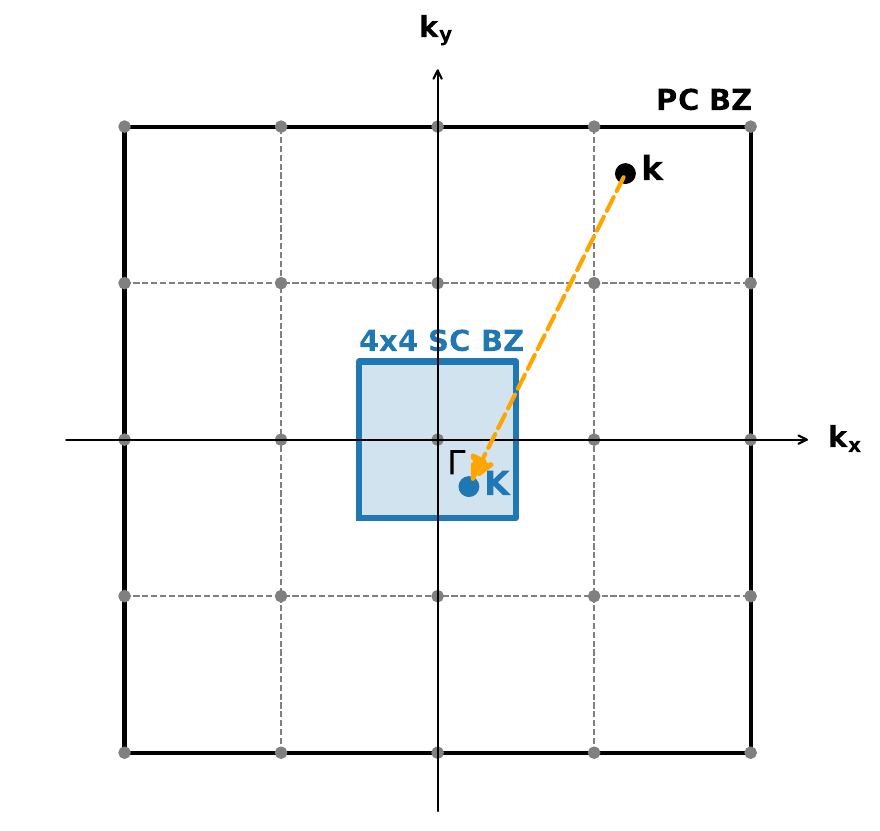}
    \caption{Illustration of a two-dimensional square PC BZ and $4 \times 4$ SC BZ, together with a $\mbf{k}$ wavevector in the PC and its corresponding folded $\mbf{K}$ in the SC BZ. Commensurate $\mbf{k}$ are those of the vertices (gray dots), which fold to $\Gamma$, and correspond to reciprocal lattice vectors of the SC. The edge of the BZ along $x$ is as usual just $1/2 \mbf{b}_1$, while for the SC BZ it is $1/2\tilde{\mbf{b}}_1$.} 
    \label{fig:BZ}
\end{figure}

We will compute the ground state of the distorted configurations using norm conserving pseudopotentials and a plane wave basis in ABINIT~\cite{verstraete25}. Formally, to compute the inner product, we consider the region of space defined by the SC, where both $|n\mbf{k}\rangle$ and $|J\rangle^I$ in Eq.~\eqref{eq:Ginit} satisfy the periodicity conditions.
If the $\mbf{k}$ of interest is not commensurate with the SC, we use a wavevector grid $\mbf{K}$ in the SC to fold $\mbf{k}$ correctly to a point in the SC Brillouin Zone (BZ), extending the Born-von K\'arm\'an volume accordingly to cover the full
$\mbf{K}$-grid. Throughout, all equations are written for $\mbf{K}=0$ for simplicity; the
general case is illustrated in Fig.~\ref{fig:BZ}.
The undistorted eigenfunctions are

\be
\psi_{n\mbf{k}} = \f{1}{\sqrt{\mathcal{V}}}\sum_\mbf{G} c_{n,\mbf{k}+\mbf{G}} e^{i(\mbf{k}+\mbf{G}) \cdot \mbf{r}},
\ee

\ni where $\mathcal{V}$ is the volume of the SC, and $\mbf{G}$ are the PC reciprocal lattice with basis $\{\mbf{b}_i\}$. Similarly, the distorted eigenfunctions are

\be
\psi_J = \f{1}{\sqrt{V}} \sum_\mbf{g} \tilde{c}_{J,\mbf{g}} e^{i \mbf{g} \cdot \mbf{r}},
\ee

\ni where $\mbf{g}$ are the reciprocal lattice vectors of the SC.  Since we will be using SCs of size $N_s \times N_s \times N_s$, $\mbf{g}= \sum_{i=1,3} m_i \tilde{\mbf{b}}_i$, with $m_i$ integers and $\tilde{\mbf{b}}_i=\mbf{b}_i/N_{s}$. Wavefunctions are normalized to 1, and planewaves are orthogonal, so

\be
\langle n \mbf{k} | J \rangle = \sum_\mbf{G} c^\ast_{n,\mbf{k}+\mbf{G}} \tilde{c}_{J,\mbf{g}=\mbf{k}+\mbf{G}} .
\label{eq:overlap}
\ee

\ni The sum is over $\mbf{G}$ since the set $\{\mbf{k}+\mbf{G}\}$ is contained in $\{\mbf{g}\}$, the set of SC reciprocal lattice vectors.

\subsection{Perturbation theory}

The AHC self-energy is given by the lowest order diagrams

\be
\Sigma_{\mbf{k}n}(\o) = \Sigma^\mathrm{Fan}_{\mbf{k}n}(\o) + \Sigma^\mathrm{DW}_{\mbf{k}n},
\label{eq:Sigma_AHC}
\ee

\ni where

\be
\begin{split}
\Sigma^\mathrm{Fan}_{\mbf{k}n}(\o) = \f{1}{N_\mbf{q}}\sum_{\mbf{q}s,m} |g^{\mbf{q}s}_{\mbf{k}nm}|^2 & \left( \f{n_{\mbf{q}s} + f_{\mbf{k}+\mbf{q}m} }{\o - \vare_{\mbf{k}+\mbf{q}m} + \o_{\mbf{q}s} + i \d} \right. \\
& \left. +  \f{n_{\mbf{q}s} + 1 - f_{\mbf{k}+\mbf{q}m} }{\o - \vare_{\mbf{k}+\mbf{q}m} - \o_{\mbf{q}s} + i \d} \right),
\end{split}
\label{eq:Fan}
\ee

\ni and

\be
\Sigma^\mathrm{DW}_{\mbf{k}n} = \f{1}{N_\mbf{q}} \sum_{\mbf{q}s} g^\mathrm{DW,\mbf{q}s}_{\mbf{k}nn}(2 n_{\mbf{q}s}+1),
\ee

\ni with $g^{\mbf{q}s}_{\mbf{k}nm}$ the first order electron-phonon matrix elements, $g^{\mathrm{DW},\mbf{q}s}_{\mbf{k}nn}$
the diagonal second order electron-phonon matrix elements~\cite{Giustino2017}, $f_{\mbf{k}m}$ the Fermi-Dirac occupation factor, and $N_\mbf{q}$ the number of $\mbf{q}$ vectors, or equivalently, the size of the SC. 
Notice that this is a diagonal approximation in the band indices. In the adiabatic approximation, which corresponds to removing the phonon frequencies from the denominators,

\be
\Sigma^\mathrm{Fan}_{\mbf{k}n}(\o) = \f{1}{N_\mbf{q}}\sum_{\mbf{q}s,m} |g^{\mbf{q}s}_{\mbf{k}nm}|^2 \left( \f{2 n_{\mbf{q}s} + 1}{\o - \vare_{\mbf{k}+\mbf{q}m}+ i \d} \right).
\label{eq:Fan_ad}
\ee

\ni In the weak-coupling limit, the AHC approach can be expected to coincide with the NPG result. However, as we will shortly see, strictly truncating the perturbation series at lowest order can yield results that are manifestly unphysical.

\subsection{Limitations of perturbation theory}
\label{sec:LimsAHC}

\subsubsection{Overview}

As mentioned in the Introduction, AHC has several limitations: when applied via the Dyson equation, it incorrectly captures the distance between the QP peak and satellite, and yields an inaccurate QP position; it also neglects anharmonic effects, and introduces errors due to the rigid ion approximation in existing implementations.

The issue of introducing AHC into the Dyson equation is one of self-consistency. The position of the satellites is referred to bare energies as opposed to the QP peak~\cite{Nery2018,Abreu2022}. Also, as demonstrated in the context of the Fröhlich model~\cite{Nery2018}, evaluating the self-energy at the bare energy ($\omega = \varepsilon_{\mathbf{k}n}$, also known as ``on-shell") yields a QP shift much closer to the exact result than extracting the pole from the Dyson-AHC equation. This works better because the denominator is then a difference of two bare
energies, $\vare_{\mbf{k}n} - \vare_{\mbf{k}+\mbf{q}m}$. 
This consistency is broken in Dyson-AHC: the QP peak $\omega_\mathrm{QP}$ is not located at the bare value, so the approximate lineshift $\Sigma_{\mbf{k}n}(\omega_\mathrm{QP})$, mixing
the renormalized $\o_\mathrm{QP}$ with bare intermediate energies, differs from the on-shell $\Sigma_{\mbf{k}n}(\omega=\varepsilon_{\mathbf{k}n})$. Alternatively, a self-consistent~\cite{Lihm2024} approach resolves this by ensuring internal lines are also updated with the renormalized energy. NPG inherently preserves this structural consistency and further avoids an arbitrary diagrammatic truncation.

We focus here on an underappreciated limitation which arises from Dyson-AHC:
its failure to describe spectral weight at the band edges (the top of the
valence band, VBM, and the bottom of the conduction band, CBM) at $T=0$, and it remains a
poor approximation at higher temperatures.
The single-particle Green's function physically describes the propagation of an added bare excitation: either injecting an electron into the conduction band or extracting one (creating a hole) from the valence band. Because this newly introduced carrier interacts with the lattice, it does not correspond to an exact eigenstate of the full electron-phonon Hamiltonian, resulting in a probability distribution $A(\o)$ of finding the system at some energy $\o$. 
In the Fr\"ohlich model featuring dispersionless optical phonons, the QP peak
at the absolute band edge at $T=0$ is in fact a perfect Dirac delta
function~\cite{Mishchenko2000}. However, broadening does appear in the
satellite region, separated from the QP peak by the optical phonon frequency.
With acoustic modes, by contrast, the vanishing $\omega_{\mbf{q}s}\to0$ as $\mbf{q}\to0$
in the denominator of Eq.~\eqref{eq:Fan} allows spectral weight to appear
immediately adjacent to the QP,
and the imaginary part of the Green's function will not be zero in
the region neighboring the QP peak.

Let us show in more detail what occurs at the top of the valence band at $T=0$ using AHC. Eq.~\eqref{eq:Fan} reduces to

\be
\Sigma^\mathrm{Fan}_{\mbf{k}n}(\o) = \f{1}{N_\mbf{q}}\sum_{m\mbf{q}s} |g^{\mbf{q}s}_{\mbf{k}nm}|^2 \f{1}{\o - \vare_{ \mbf{k}+\mbf{q}m} + \o_{\mbf{q}s} + i \d}.
\ee

\ni In the typical case, the gap narrows upon including zero-point renormalization and temperature, and the valence band gets shifted upwards. Then, $\o_{QP} - \vare_{\mbf{k}+\mbf{q}m} + \o_{\mbf{q}s} > 0 $ for any phonon frequency and valence band $m$,  and $\Im \mathrm{m} \Sigma(\o) = 0$ in the region of the QP peak. When $\Im\mathrm{m} \Sigma(\o)=0$, $i \d$ in Eq.~\eqref{eq:Sigma} cannot be discarded as in Eq.~\eqref{eq:final}. In the diagonal approximation, $G_\mbf{k}$ has a pole at $\o_\mathrm{QP}  - \vare_{\mbf{k}n} - \Re\mathrm{e}\Sigma_{\mbf{k}n}(\o_\mathrm{QP}) = 0$ and the spectral function Eq.~\eqref{eq:A} yields a delta function at
$\o_{\mathrm{QP}}$, while the remaining structure stays referred to the
bare energies rather than to the renormalized ones.

\subsubsection{Vanishing of $\Im\mathrm{m}\Sigma$ inside the gap to all orders}

More generally, the perturbative self-energy has no imaginary part inside
the bare gap at $T = 0$, at any order. The physical content is simple: at $T=0$ nothing is available that lowers the intermediate-state energy $\Omega_\chi$, since $|\Phi_0\rangle$ has no
phonons to absorb, no conduction electrons to remove and no valence holes
to fill. Electron-number conservation means any further electron must be
accompanied by a hole, raising $\Omega_\chi$ by at least
$E_{\mathrm{gap}}$; and each phonon can only be created, raising
$\Omega_\chi$ by its frequency. The pole sits at $\o = +\Omega_\chi$ when
an electron is added and at $\o = -\Omega_\chi$ when one is removed, so
both effects displace it away from the gap -- upward from the conduction
edge, downward from the valence edge -- and none can land inside, whatever
the order of the diagram. The full proof is given in
Appendix~\ref{sec:zero_temperature}, and its structure is the following.

By the Gell-Mann--Low theorem, the time-ordered Green's function is
referred to the noninteracting reference state $|\Phi_0\rangle$, the
filled valence determinant times the phonon vacuum, and the $S$ matrix is expanded in powers of the
electron-phonon potential. Within each chronological ordering of the
vertices the operators are strictly ordered, so that inserting complete
sets of eigenstates of $H_0$ between them assigns to each interval a
single phase $e^{-i(E_\chi - E_0)u}$, with $|\chi\rangle$ the
intermediate state, $E_0$ the energy of $|\Phi_0\rangle$, and $u$ the
length of the interval. The interval lengths are independent variables,
so the time integrals factorize and yield one energy denominator per
intermediate state, the remaining factors being independent of $\o$.
Because the electron-phonon potential conserves the electron number,
every $|\chi\rangle$ that appears lies in the sector with one electron
more, or one fewer, than $|\Phi_0\rangle$, and the states of those
sectors are the filled valence bands with $p$ electrons added in
conduction states, $h$ removed from valence states and any number of
phonons, with $p - h = 1$ ($h - p = 1$ for removal).

Summing the energies of such a configuration, an energy denominator
vanishes at

\be
\o = \vare_{m\mbf{k}'} \pm \o_{\mbf{q}_1 s_1} \pm \dots \pm
\o_{\mbf{q}_L s_L} + \sum_i \left( \vare_{a_i} - \vare_{b_i} \right),
\label{eq:poleform}
\ee

\ni where $m\mbf{k}'$, $a_i$ and $b_i$ label intermediate electronic
states. The terms are of three kinds: one unpaired band energy, phonon
frequencies, and differences of pairs.
That the electronic energies enter only in this way, with signs summing to
one, follows from electron-number conservation. The phonon signs are not
constrained in general, since a phonon may be either emitted or absorbed,
but absorption requires a phonon to be present. At $T=0$ there are none,
so only emission occurs, and every $E_\chi - E_0$ contains the phonon
energies with a positive sign. The two sectors differ in that the pole
lies at $\o = E_\chi - E_0$ for addition and at $\o = -(E_\chi - E_0)$ for
removal, so the phonon frequencies appear all with a positive sign for the
states reached by adding an electron and a negative one for those reached
by removing one. Differences of band energies are absent at the
one-phonon-line level and first arise in the crossed diagram with two
phonon lines.

The energy of any such configuration therefore satisfies $\o \geq
\vare_{\mathrm{CBM}} + h\, E_{\mathrm{gap}}$ for addition and $\o \leq
\vare_{\mathrm{VBM}} - p\, E_{\mathrm{gap}}$ for removal, so that no
energy denominator can vanish inside the gap. The purely electronic
combinations, which require $h \geq 1$ for addition (resp.\ $p \geq 1$ for
removal), are confined to at least $E_{\mathrm{gap}}$ beyond the band
edges. A perturbative spectral function therefore cannot acquire spectral
weight inside the gap at any order, beyond the sharp QP peak shifted by
$\Re\mathrm{e}\,\Sigma$: PT can shift the band edge but cannot give it a
finite width. At finite temperature phonons are present,
absorption is allowed and all the sign combinations of
Eq.~\eqref{eq:poleform} occur, so that in-gap weight becomes exponentially
small rather than forbidden, suppressed by the Fermi-Dirac and
Bose-Einstein factors that weight each position, e.g.\
$(1-f_{v\mbf{k}'}) + n_{\mbf{q}s}$ for the Fan term at $\o =
\vare_{v\mbf{k}'} + \o_{\mbf{q}s}$
(Appendix~\ref{sec:finite_temperature}). Both the pole positions and these
occupations remain referred to the unperturbed energies, however, so the
finite-temperature description is no better founded than the $T=0$ one; it
merely fails less conspicuously.

NPG does not suffer the issue of failing to account for spectral weight
inside the bare gap. Each distortion yields a different eigenstate of the
Hamiltonian and thus a different pole in the Green's function, and these
poles are those of the distorted system rather than of the unperturbed one.
Integrating over the nuclear distribution therefore places weight wherever
the distorted eigenvalues lie, inside the bare gap included, independently
of whether the gap opens or closes on average (in most materials it typically closes).

\subsubsection{Toy model}

\vspace{0.5cm}
To better understand the differences between AHC and NPG, let us consider a one-band toy model, with coupling 

\be
V(u) = g u + \tfrac{1}{2} g' u^2,
\label{eq:toy_model}
\ee

\ni where $g$ and $g'$ are the (intraband) first and second order electron-phonon matrix elements, respectively. The displacement $u$ follows the probability distribution $P(u)$ just as in Eq.~\eqref{eq:proba}, but for one phonon mode,

\begin{equation}
P(u) = \frac{1}{\sigma\sqrt{2\pi}} \exp\left(-\frac{u^2}{2\sigma^2}\right),
\end{equation}

\ni with $\s$ the root mean square displacement of the phonon. We analyze this example using AHC, NPG, and also a self-consistent approach and the cumulant method.

\paragraph{NPG}

The distorted configurations $I$ can be labeled now by the continuous index $u$. Let us start by first omitting the DW term. Let $x=\o+i\d-\vare_0$. From Eqs.~\eqref{eq:GI} and \eqref{eq:average} we have

\be
G = \int du \f{P(u)}{x - g\, u},
\label{eq:G}
\ee

\ni and taking the imaginary part yields
 
\be
A(\omega) = \frac{1}{\sqrt{2\pi \a}} \exp\left(-\frac{(\omega - \varepsilon_0)^2}{2\a}\right),
\label{eq:A_intra_NP}
\ee

\ni which does have a width, with $\a \equiv g^2\s^2$. The spectral function is a Gaussian, a completely different form from that obtained with AHC. 

Adding back the DW term, the spectral function becomes

\begin{equation}
A(\omega) = \int du \, P(u) \, \delta\!\left(\omega - \varepsilon_0 - gu - \tfrac{1}{2}g'u^2\right),
\end{equation}

\ni which is simply the probability distribution of the pole position $\omega_p(u) = \vare_0+gu+\tfrac{1}{2}g'u^2$ under thermal fluctuations. Since $P(u)$ is Gaussian with standard deviation  $\sigma$, the pole position has simply a mean $\varepsilon_0 + \alpha'$, with $\a' \equiv g'\s^2/2$ (the DW shift), and calculating the standard deviation gives $\sqrt{\a + 2 \a'^2}$. 
It may seem surprising that DW enters into the width. In AHC, the width
comes from the imaginary part of the Fan term: DW is real, so it does not
contribute. This is consistent at leading order, since $\a$ is of order 2
in the displacements while $\a'^2$ is of order 4, so that in the weak
coupling case the width reduces to $\sqrt{\a}$, independent of DW. At
higher order, however, two DW vertices appear in a diagram analogous to
Fan, but with two phonon lines instead of one (since each DW vertex carries
two phonon legs), and this diagram does have an imaginary part, so DW also
enters the width at higher order in PT.

\paragraph{AHC}

Let us briefly review the connection between NPG and AHC. One can formally write Eq.~\eqref{eq:GI}

\be
\begin{split}
G^I & = \f{1}{G_0^{-1} - V^I} = \f{1}{1 - G_0 V^I} G_0 \\
     & = G_0 + G_0 V^I G_0 + G_0 (V^I G_0)^2 + \cdots .
\end{split}
\ee

\ni Performing the thermal average term by term, and using
$\langle u^{2k}\rangle = (2k-1)!!\,\s^{2k}$ (the double factorial includes only odd or even terms), yields (again omitting DW)

\be
 G  = \f{1}{x}\sum_{k=0}^{\infty}
    (2k-1)!!\left(\f{\a}{x^2}\right)^{k},
\label{eq:G_avg_series}
\ee

\ni where $(2k-1)!! = 1\cdot 3 \cdots (2k-1)$. The $k=1$ term coincides with the first-order term of the AHC
Green's function, and the series reproduces the full
perturbative set of diagrams with a single electron propagator, i.e.\ every phonon topology at every order.
 This series does not converge: the coefficients $(2k-1)!!$ grow
factorially, so it is the asymptotic expansion of
Eq.~\eqref{eq:A_intra_NP} rather than a convergent representation of it.
This is the generic situation for perturbative expansions in quantum field
theory and many-body physics~\cite{Dyson1952,Lipatov1977}, where the
coefficients typically grow factorially because the number of diagrams at
a given order does. Divergence of the full series does not invalidate
low-order truncations, however: truncating after $N$ terms leaves an error
of the order of the first term omitted, so the first few orders remain
controlled as the expansion parameter goes to zero. What fails is the limit
of many terms at fixed coupling. Here the ratio of consecutive terms is
$(2k+1)\a/x^2$, so they decrease only while $k \lesssim x^2/2\a$. Note also that the
relevant parameter is not the coupling alone but $\a/x^2$, and the
expansion therefore degrades as $\o$ approaches $\vare_0$.

At the level of Eq.~\eqref{eq:GI}, before thermal averaging, the condition for
the leading terms to dominate is simply $|G_0 V| \ll 1$. Adding back the DW term, this gives heuristically, for the dominant displacement $u \sim \s$, the condition
$|\o - \vare_0| \gg \sqrt{\a} + \a'$.

The AHC self-energy is simply $\Sigma=\tfrac{\a}{x}+\a'$ and the Green's function is

\be
G = \f{1}{x - \tfrac{\a}{x}-\a'},
\label{eq:G_intra_PT}
\ee

\begin{figure}
    \centering
    \includegraphics[width=\linewidth]{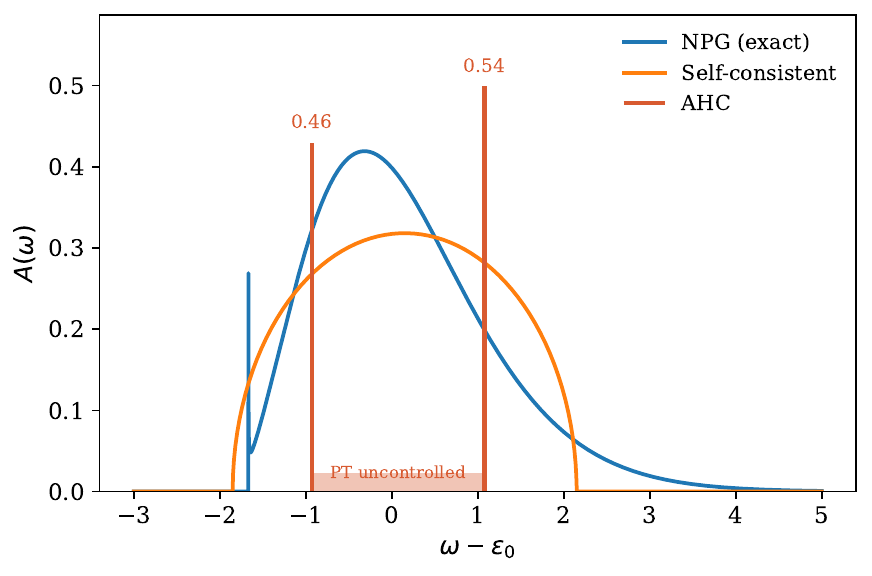}
    \caption{Spectral function of the toy model of Eq.~\eqref{eq:toy_model} for AHC, NPG and SCBA. The AHC spectral function completely misses the width given by NPG. The weight of each pole is indicated next to it. SCBA has a FWHM almost 50\% larger than NPG without the DW term, and the shape is very different.}
    \label{fig:toy_model}
\end{figure}

\ni whose poles lie at $\o = \vare_0 + \tfrac{1}{2}\big(\a' \pm
\sqrt{\a'^2 + 4\a}\big)$, reducing to $\vare_0 \pm \sqrt{\a}$ when the DW
term is dropped. So PT is an uncontrolled approximation precisely between the poles, where most of the spectral weight of Eq.~\eqref{eq:A_intra_NP} lies. See Fig.~\ref{fig:toy_model}.

\paragraph{Self-consistency}

Another approach that resums an infinite subset of irreducible diagrams is the self-consistent Born approximation (SCBA), which consists in replacing the bare $G_0$ with the dressed $G$ in the electronic line of the Fan self-energy. The SCBA also recovers a finite width at the band edge. In our toy model, the self-consistent Fan plus DW self-energy reads

\begin{equation}
\Sigma(\omega) = \frac{\a}{x - \Sigma(\omega)} + \a',
\end{equation}

\ni which reduces to a quadratic equation with solution

\begin{equation}
\Sigma(\omega) = \alpha' + \frac{(x - \alpha') - \sqrt{(x - \alpha')^2 - 4\alpha}}{2},
\end{equation}

\ni where for negative values of $z$, we take $\sqrt{z}=+i\sqrt{-z}$. In this way $\Im\mathrm{m} \Sigma < 0$ (retarded self-energy).

The self-energy develops a branch cut when $|x - \alpha'| < 2 \sqrt{\a}$, giving a finite imaginary part

\begin{equation}
\Im\mathrm{m}\,\Sigma(\omega) = -\frac{1}{2}\sqrt{4 \a - (x - \alpha')^2}.
\end{equation}

\ni This leads to the spectral function

\begin{equation}
A(\omega) = \frac{\sqrt{4\a - (\omega - \varepsilon_0 - \alpha')^2}}{2\pi \a}.
\label{eq:A_SC}
\end{equation}

\ni This is a semicircle centered at $\varepsilon_0 + \a'$ with radius $2\sqrt{\a}$.
The full width at half maximum (FWHM) is $2 \sqrt{3\a}$, which is about 47\% larger than the FWHM $2 \sqrt{2 \ln 2}\,\sqrt{\a}$ of the Gaussian NPG spectral function with no DW term. 

\paragraph{Cumulant}

Another approach that goes beyond simple PT is the cumulant expansion, which
writes the Green's function in the time domain as an exponential,
$G(t) \propto e^{-i\vare_0 t}e^{C(t)}$, with $C(t)$ built from the lowest-order
self-energy. The exponentiation generates higher-order diagrams in an
approximate way, and is known to become exact for bands and electron-phonon matrix elements that do not depend on $\mbf{k}$ (the independent
boson model)~\cite{Dunn1975}, which is the case of the present model. Indeed,
as shown in Appendix~\ref{sec:PT_vs_NP_vs_SC}, the cumulant reproduces the
exact NPG spectral function, Eq.~\eqref{eq:A_intra_NP}.

\vspace{0.5cm}
The comparison between AHC, NPG and SCBA can be visualized in Fig.~\ref{fig:toy_model}. AHC completely fails to describe the spectral function in this toy model, giving just two delta functions as opposed to finite values in an extended region as in NPG. SCBA is much better than AHC, but the FWHM differs about 50\% relative to NPG for weak coupling and does not depend on DW, and the shape is drastically different.

\subsection{Degeneracy}
\label{sec:degeneracy_main}

In principle, the self-energy has off-diagonal components in the band indices that are not 0, and this full self-energy is needed to determine $G$ through Eq.~\eqref{eq:Sigma}. However, if bands are not close to each other, it is generally expected that such elements will be small, and the diagonal approximation is frequently used. What happens however if energies are degenerate?

The self-energy inherits the full symmetry of the Hamiltonian, so it must
commute with every symmetry operation of the crystal that leaves $\mathbf{k}$
invariant. At $\Gamma$, this is the entire point group.
Grouping bands by their irreducible representation, this commutation makes $\Sigma_\Gamma$ block-diagonal: it
cannot connect states of different symmetry. For the threefold-degenerate
valence states ($\Gamma_{25'}$), Schur's lemma forces $\Sigma$ to act as a
single scalar times the identity on the degenerate manifold, so the
off-diagonal elements among these states vanish exactly by symmetry. (They could still couple to other bands of the same
irreducible representation, but the resulting admixture is suppressed by the large energy
separation, so the VBM self-energy is diagonal to excellent accuracy.)
An exception could arise at accidental degeneracies, where no symmetry
constrains the off-diagonal self-energy and it need not vanish. However, bands
that cross accidentally are typically of different orbital character, which
tends to suppress the off-diagonal electron-phonon matrix element between them.
The net importance of such crossings is therefore case-dependent and cannot be
settled in general.

Away from $\Gamma$, however, the little group is smaller (and typically trivial), and
off-diagonal self-energy elements between quasi-degenerate states are no longer
forced to vanish. These could affect the spectral function and, in particular,
the effective mass. The effect is expected to be most significant in polar
materials, where the Fr\"ohlich coupling $|g| \sim 1/q$ enhances the
$\mathbf{q} \to 0$ region near the band extremum. While degeneracies in the
presence of Fr\"ohlich coupling have been examined in Ref.~\cite{Guster2021},
that analysis is restricted to the lowest-order perturbative terms. Among
non-perturbative approaches, the diagrammatic Monte Carlo method of
Ref.~\cite{Mishchenko2000} treats the single-band Fr\"ohlich model, in which
degeneracy is absent by construction, while the first-principles diagrammatic
Monte Carlo method of Ref.~\cite{Luo2025}, applied to real multi-band materials,
includes such effects in principle, but the specific contribution from
quasi-degenerate bands has not been isolated. It therefore warrants a dedicated study beyond AHC.
Although NPG includes off-diagonal contributions, it would require an impractically large SC
to study them from first-principles.

\begin{figure}[t]
\begin{tabular}{cc}
\includegraphics[scale=0.42]{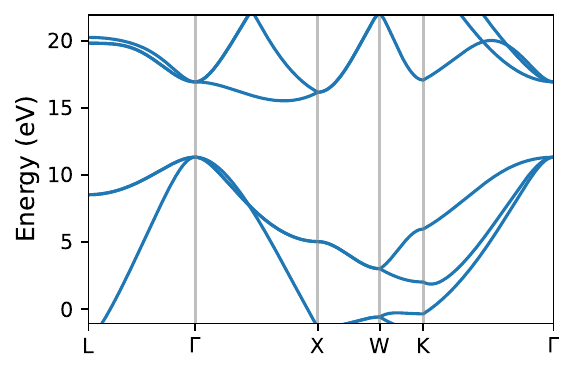} &
\includegraphics[scale=0.42]{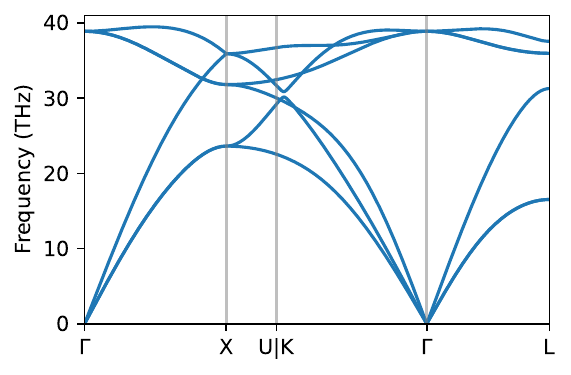} \\
(a) & (b) \\
\end{tabular}
\caption{(a) Band structure of diamond. We focus on the spectral function at $\Gamma$, around the VBM and CB$_\Gamma$, and also study the spectral function at the CBM (along $\Gamma$--$X$) and at the corresponding $\mbf{k}$-point in the valence region. (b) Phonon dispersion of diamond which enters into the probability distribution Eq.~\eqref{eq:proba}, determining the distorted configurations of NPG and the AHC self-energy Eq.~\eqref{eq:Sigma_AHC}.}
\label{fig:bands}
\end{figure}

\begin{figure}[t]
\includegraphics[scale=0.45]{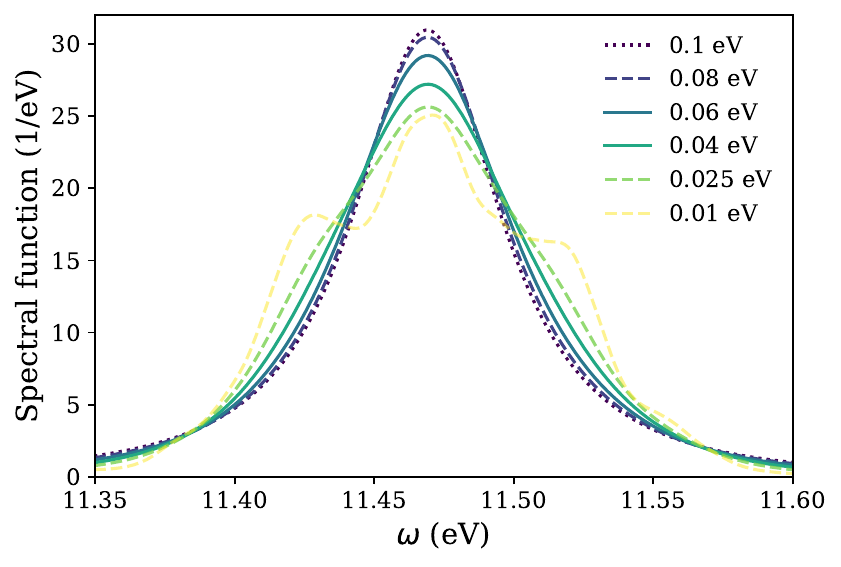} \\
(a) \\
\includegraphics[scale=0.45]{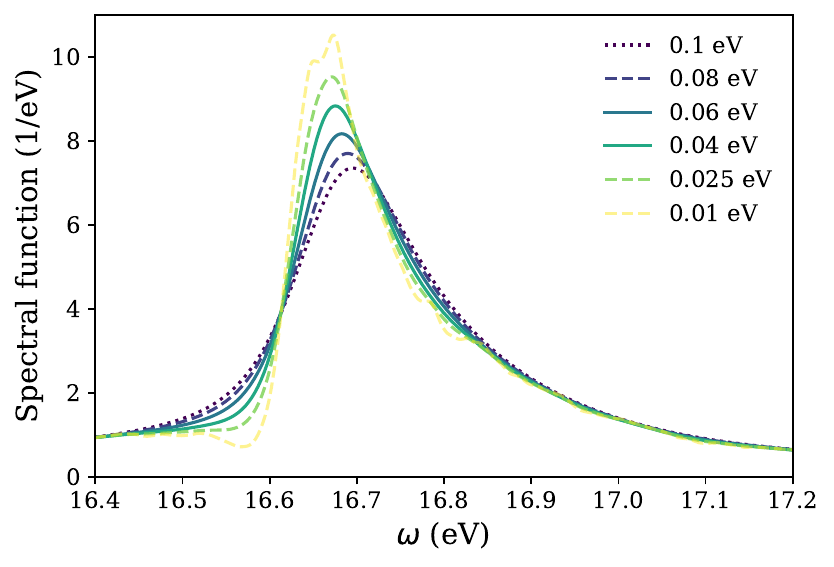} \\
(b) \\
\caption{NPG spectral functions for different values of $\d$ for $N_s=8$. (a) VBM and (b) CB$_\Gamma$. The lowest value of $\d$ that produces a smooth peak both for valence and conduction is approximately $0.04$~eV. We consider a narrow range of $\d > \d_c$ to linearly extrapolate to the limit $\d \rightarrow 0$. See Fig.~\ref{fig:extrapolation}.}
\label{fig:spectral_vs_delta}
\end{figure}

A separate issue arises at the level of the Hamiltonian. Although the
self-energy is diagonal within the degenerate manifold, the off-diagonal
electron-phonon couplings among those states are not, and they act exactly
as an intraband coupling does for an isolated band: diagonalizing the
degenerate subspace at fixed displacement turns them into diagonal shifts.
We analyze this case in detail in Appendix~\ref{sec:degeneracy}, using a
$2\times2$ toy model similar to the intraband model of the previous
section.

\section{Convergence}
\label{sec:convergence}

Since this is the first work in which NPG is applied from first-principles, we perform a detailed convergence study, before presenting full results in the next section. 
We focus on $\Gamma$, especially the VBM, but also CB$_\Gamma$.
The band structure of diamond is plotted in Fig.~\ref{fig:bands}~(a). The phonon dispersion, used to generate the distorted configurations, is shown in (b).
In the NPG formalism the ground-state energies and wavefunctions have to be calculated for each stochastic SC configuration of size $N_s$, results have to be averaged over the number of configurations $N_\mathrm{cfgs}$, and a broadening parameter $\d$ has to be chosen to make the spectral function smooth. 
In what follows, we study the convergence of NPG with $\d$, $N_\mathrm{cfgs}$ and $N_s$.

\begin{figure}[t]
\begin{tabular}{cc}
\includegraphics[scale=0.38]{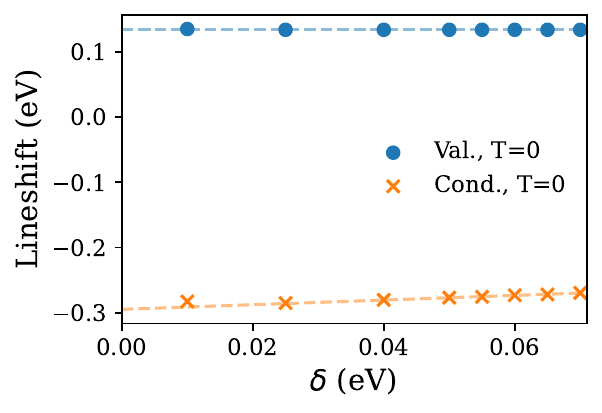} &
\includegraphics[scale=0.38]{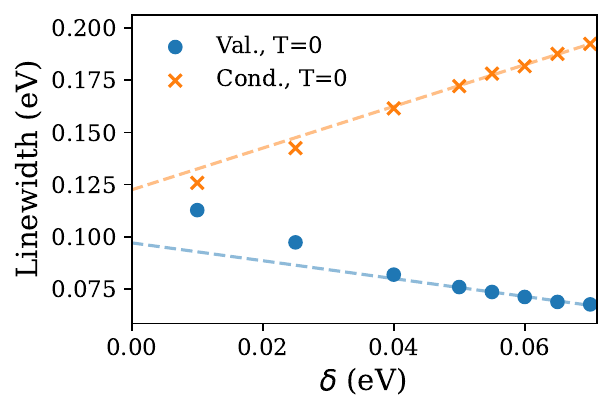} \\
(a) & (b) \\
\end{tabular}
\caption{(a) Lineshift and (b) linewidth at the VBM and CB$_\Gamma$ as a function of $\d$. To obtain the linewidth at $\d=0$ we use a linear extrapolation on the 0.05--0.07~eV range.}
\label{fig:extrapolation}
\end{figure}

We first examine the evolution of the NPG spectral function with $\d$. 
Below a critical value $\d_c$, the spectral function presents spurious oscillations, which are not due to statistical under-sampling, but originate from discrete momentum sampling: distortions are only generated by momenta commensurate with the SC. 
In Fig.~\ref{fig:spectral_vs_delta} we use $N_\mathrm{cfgs}=120$. The curve with $\delta=0.025$~eV (light-green dashed line) begins to develop additional inflection points relative to the smoother curves at larger $\delta$.
Below $\d_c$ the second derivative of $A(\o)$ changes sign more than twice
within the region carrying most of the quasiparticle weight, signalling
structure that is not physical. At $\d = 0.04$~eV this no longer occurs,
though the margin is narrow. To avoid working so close to this limit, while
also avoiding values of $\d$ large enough for the linear dependence to break
down, we perform the extrapolation over the range $0.05$ to $0.07$~eV.
Peak positions remain largely insensitive to $\d$, as shown in
Fig.~\ref{fig:extrapolation}~(a). The width is not, and the extrapolation
is needed to obtain its value in the limit $\d \rightarrow 0$.

\begin{figure*}[t]
\begin{tabular}{ccc}
\includegraphics[scale=0.38]{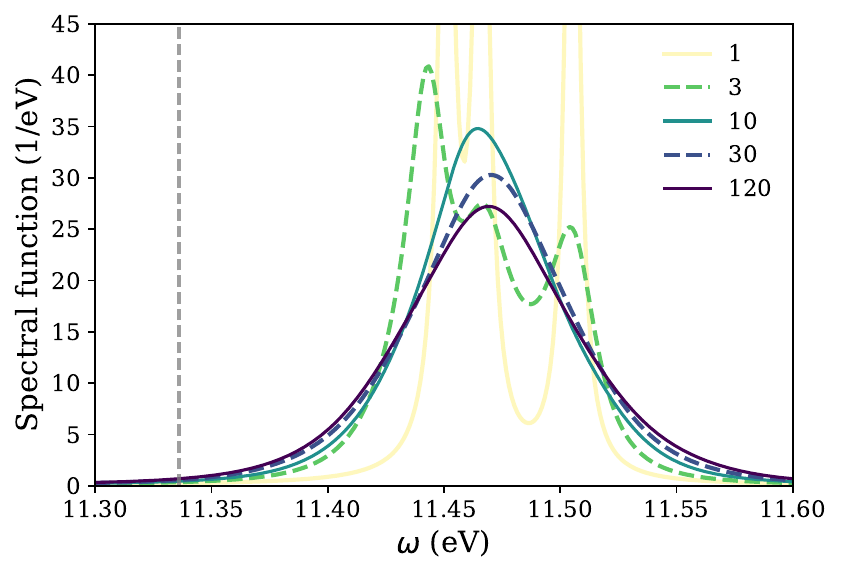} &
\includegraphics[scale=0.38]{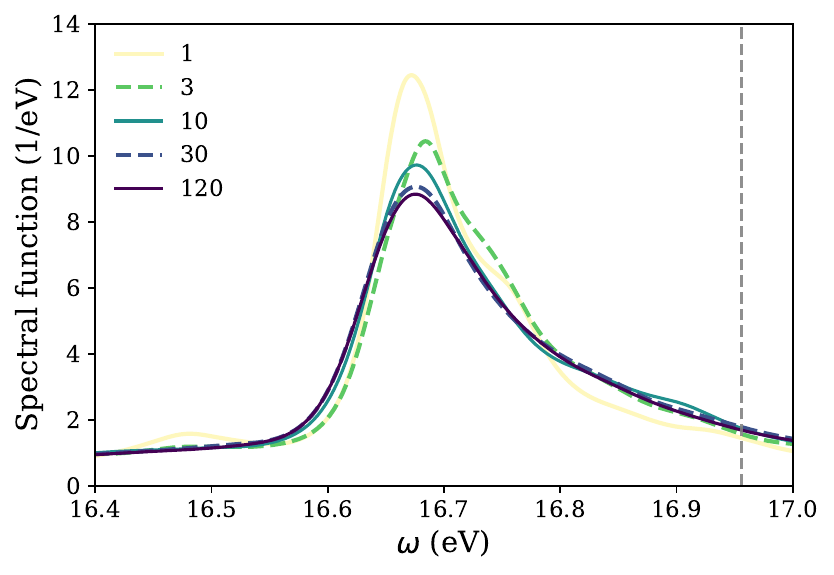} &
\includegraphics[scale=0.38]{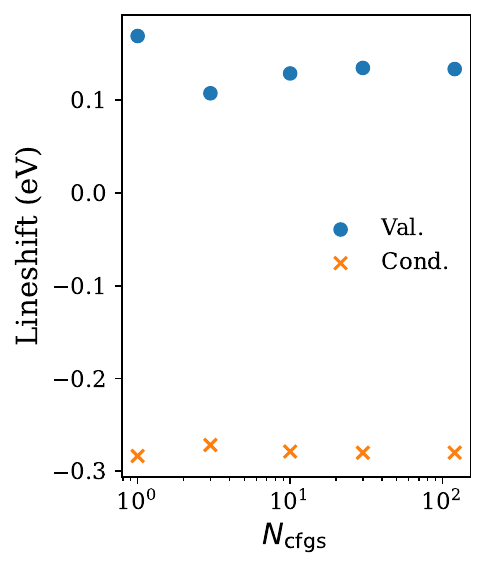} 
\includegraphics[scale=0.38]{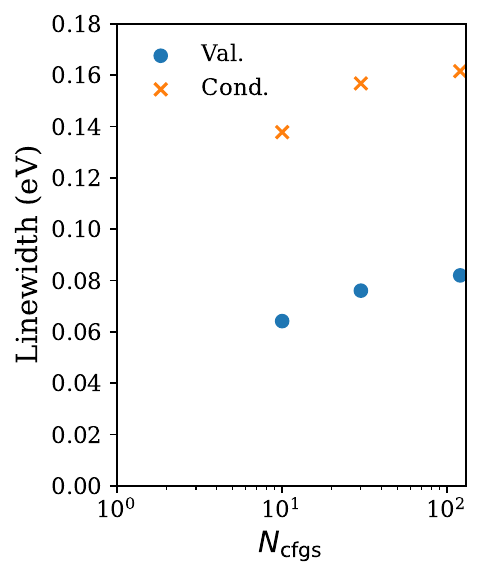}
\\
(a) & (b) & (c)
\end{tabular}
\caption{Spectral function in the (a) VBM and (b) lowest conduction band regions at $\Gamma$ averaged over different number of configurations. The vertical dashed line marks the unperturbed eigenvalue. Below 10 configurations there is not a single well defined peak (concavity changes more than two times), so the linewidth is not well defined. The extracted shift and width of the main peak are shown in (c): roughly
100 configurations are needed to converge the width to a few percent.}
\label{fig:spectral_vs_Ncfgs}
\end{figure*}

Now, using $\d=0.04$~eV, let us look at the convergence of the spectral function with $N_\mathrm{cfgs}$ (Fig.~\ref{fig:spectral_vs_Ncfgs}). It has a well-defined peak for $N_\mathrm{cfgs} \geq 10$ at the VBM, and for all $N_\mathrm{cfgs}$ values at CB$_\Gamma$. 
In both cases, the width increases slightly with $N_\mathrm{cfgs}$ (panel (c)), as larger samples progressively capture the tails of the nuclear quantum distribution. 
The width at the VBM for 30 configurations differs from 120 configurations by 7\%, and by 3\% for the CB$_\Gamma$, so results are reasonably well-converged and should differ from the exact result by at most a few meV.

\begin{figure*}[t]
\begin{tabular}{ccc}
\includegraphics[scale=0.4]{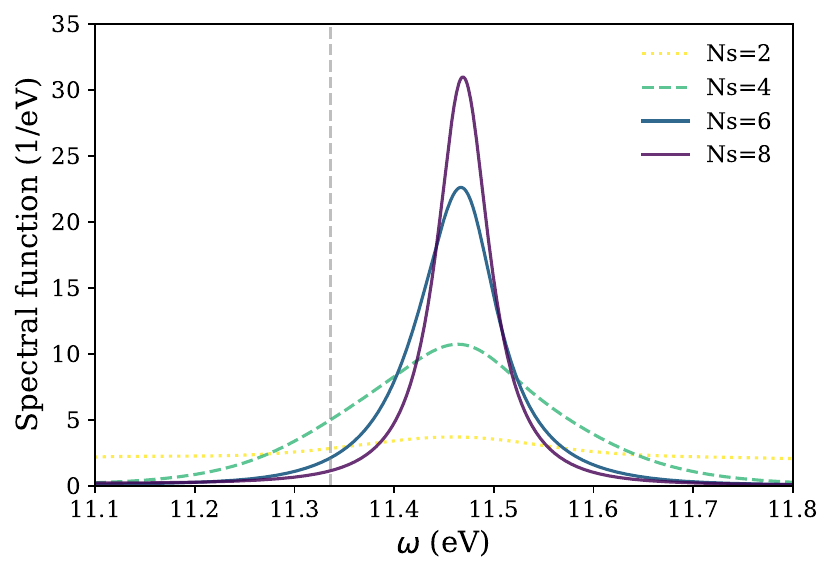} &
\includegraphics[scale=0.4]{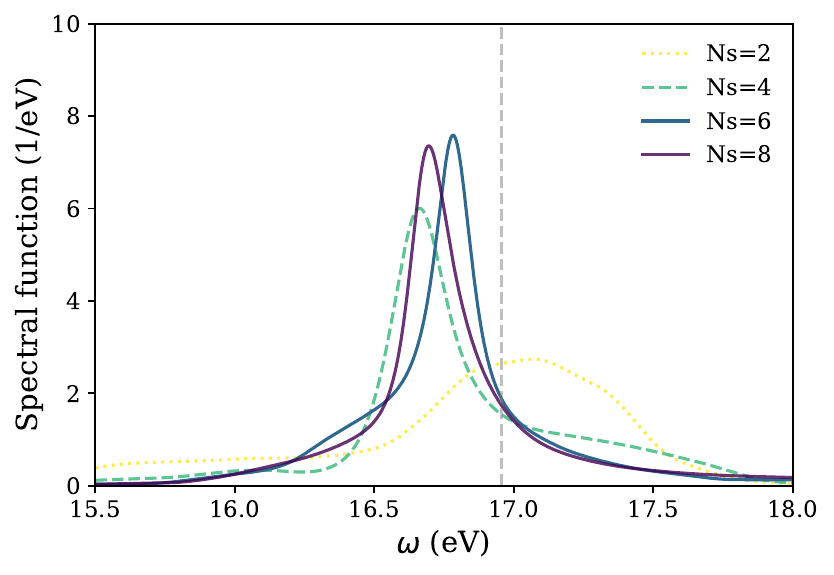} &
\includegraphics[scale=0.4]{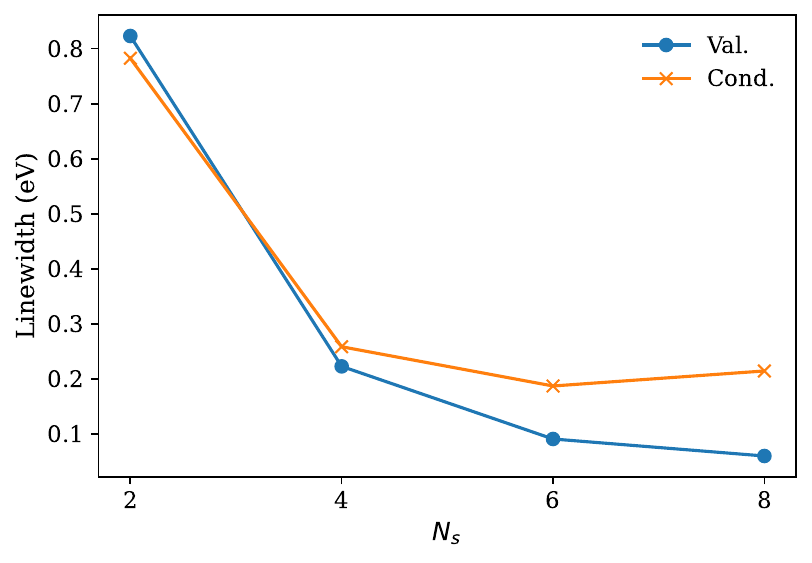} \\
(a) & (b) & (c)
\end{tabular}
\caption{Spectral function for different SC sizes at (a) VBM and (b) conduction band at $\Gamma$. Panel (c) shows the width as a function of $N_s$ in both cases. $N_s=2$ uses 300 configurations, while the other ones use 40. The linewidth
drops sharply up to $N_s = 4$ and changes much more slowly thereafter.} 
\label{fig:conv_Ns}
\end{figure*}

Smaller SCs require larger values of $\d$. We use $\d = 0.1$~eV in
Fig.~\ref{fig:conv_Ns}, which yields a smooth spectral function at $N_s=4$,
6 and 8, with a number of configurations comparable to that needed at
$\d = 0.04$~eV for $N_s = 8$.

The VBM width decreases monotonically with $N_s$, which can be attributed to the over-representation, at small $N_s$, of the intraband and degenerate effects analyzed in Appendix~\ref{sec:degeneracy}. $N_s = 2$ gives a very large broadening of around 0.8~eV.
This effect is less pronounced at CB$_\Gamma$, where the contribution from other states at similar energies leads to faster convergence.

To summarize, for diamond, the number of configurations and size of the SC are reasonably well converged at $N_\mathrm{cfgs}=120$ and $N_s=8$, while an extrapolation is required for the broadening parameter.

\medskip
It is well known that PT requires very dense $\mbf{q}$-grids to converge
the self-energy~\cite{Ponce2025,Ponce2015}, and one might expect
correspondingly large SCs to be needed for converged NPG spectral
functions. 
We find instead that SCs of order $10^3$ atoms, whose distortions are built
from the $8^3$ commensurate $\mbf{q}$ points, yield smooth spectral
functions at $\d = 0.04$~eV, whereas the AHC self-energy is evaluated on a
$64^3$ grid with $\d = 0.025$~eV. (A fully like-for-like comparison would
require the AHC calculation at the same $\d$.)
The reason is that the two methods sample different
variables, $\vare_{\mbf{k+q}m}$ against $\vare_J$. In PT, the energy denominators contain $\vare_{\mbf{k+q}m}$,
which varies over the electronic bandwidth, of order eV, as $\mbf{q}$
varies: its change between consecutive grid points must fall below the QP
width, of order tens of meV, so that grids of order $100^3$ are typically
needed. In NPG the corresponding energies are the $\vare_J$ of a
distorted SC, and these are spread over an interval of order the width
itself, since the width is precisely their spread across the ensemble.
There is no eV-scale variation to be resolved on a meV
scale, and the same resolution criterion is met with a much coarser grid.
The SC size enters through the distortion, which is a sum of
contributions from all phonons of the commensurate grid and is therefore
already a representative sample of the thermal displacements for modest
$N_s$; increasing $N_s$ refines the distortion without qualitatively
changing it, unless the phonon dispersion carries features that a coarse
grid misses entirely. What NPG does require is an average over
configurations, to turn the discrete $\vare_J$ into a smooth peak, which
partly offsets the gain of the coarse grid.

\begin{figure*}[ht!]
\begin{tabular}{ccc}
\includegraphics[scale=0.4]{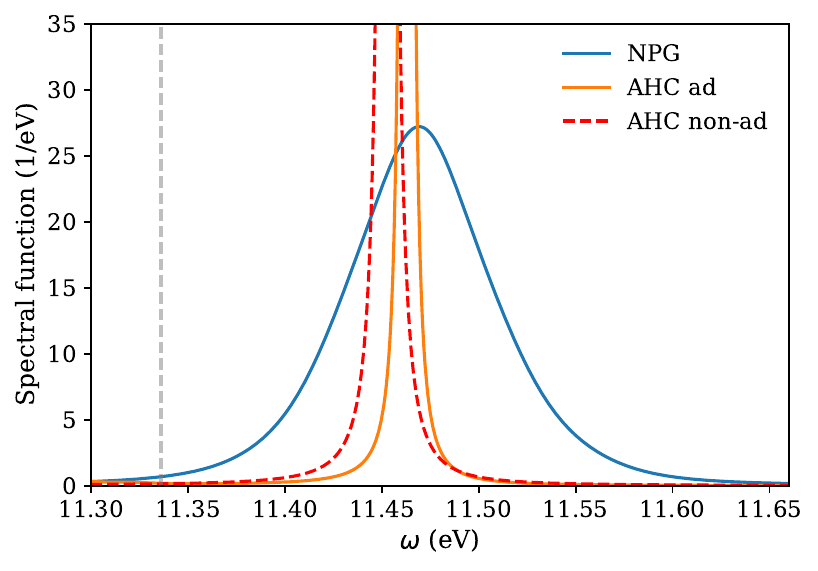} &
\includegraphics[scale=0.4]{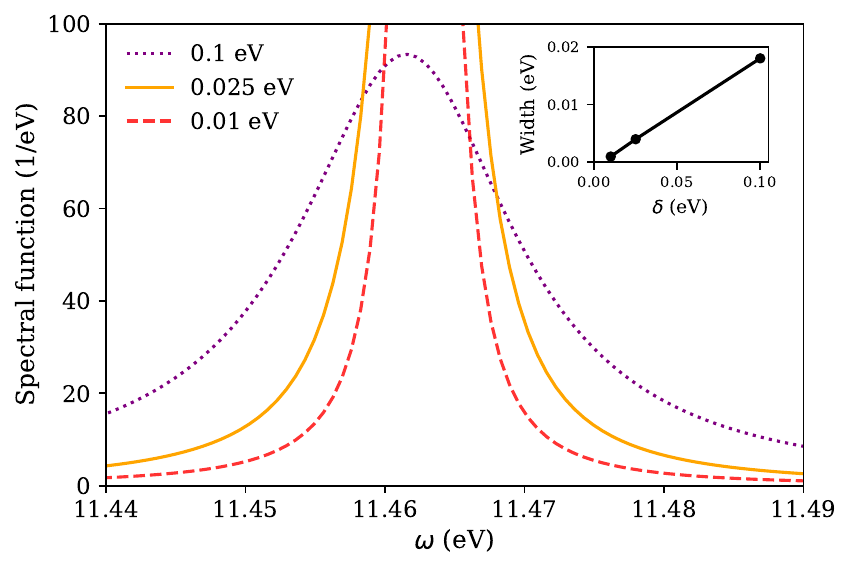} &
\includegraphics[scale=0.4]{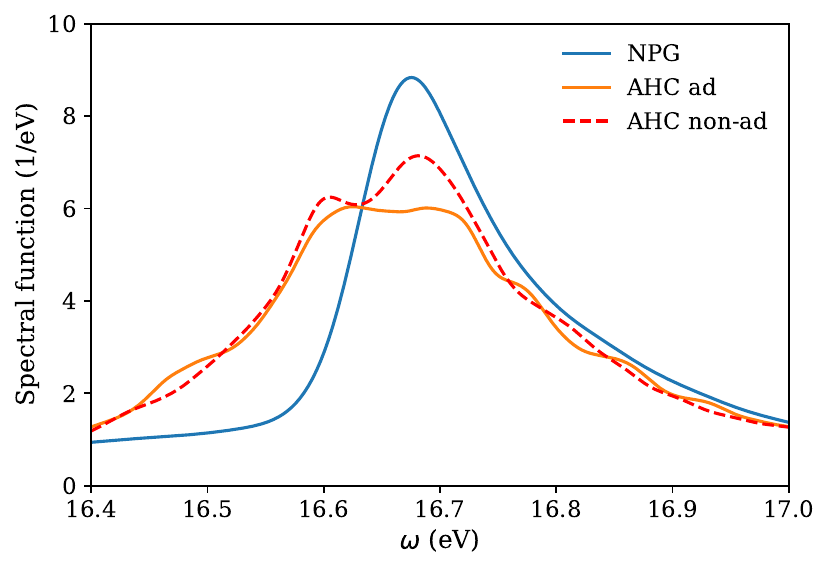} \\
(a) & (b) & (c)
\end{tabular}
\caption{Comparison between NPG, AHC adiabatic, and AHC non-adiabatic spectral functions. 
(a) VBM. While the shifts are similar, the NPG width is much larger, which is a limitation of AHC.
(b) Variation of the VBM AHC adiabatic result with the smearing width $\delta$. Inset shows a linear relation between the width and $\delta$, with 0 residual intercept: the width is purely numerical and AHC produces no weight in the gap.
(c) CB$_\Gamma$ (not the band edge). The CB$_\Gamma$ widths are closer than for the VBM, but still differ visibly, and NPG is much more asymmetric.}
\label{fig:val_NP_vs_PT}
\end{figure*}

\begin{figure}[t]
\begin{tabular}{cc}
\includegraphics[scale=0.4]{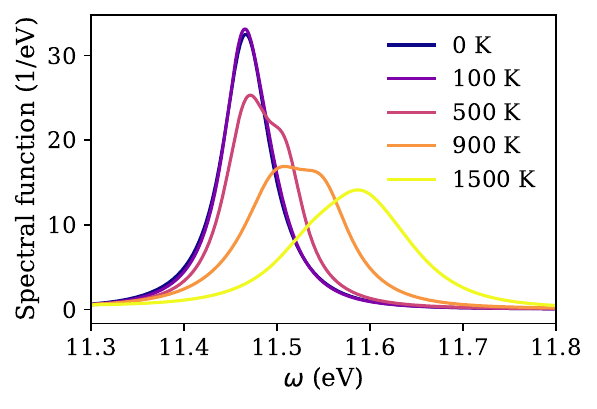} &
\includegraphics[scale=0.4]{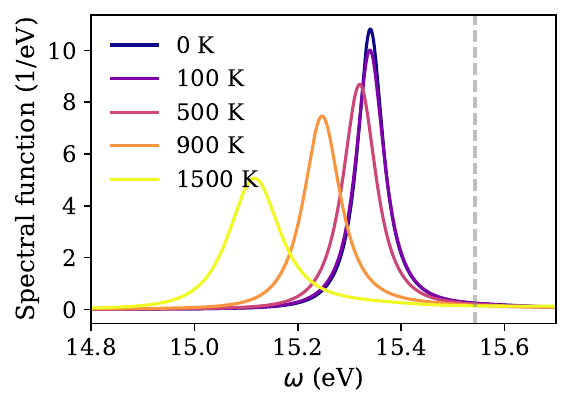} \\
(a) & (b) \\
\includegraphics[scale=0.4]{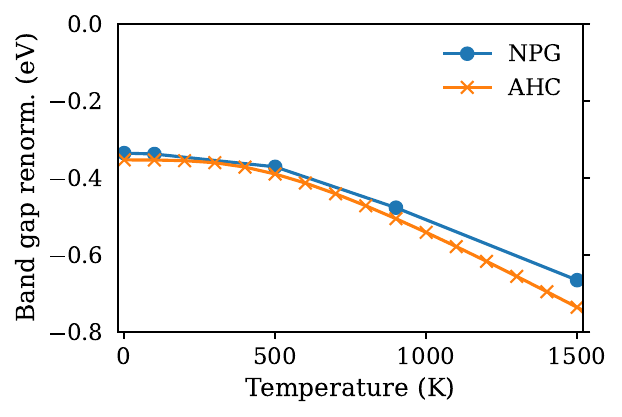} &
\includegraphics[scale=0.4]{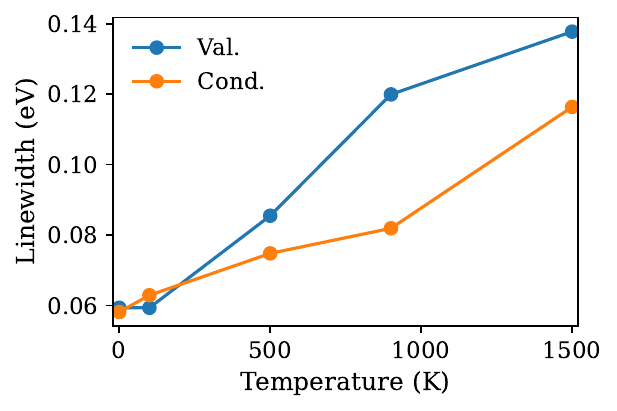} \\
(c) & (d)
\end{tabular}
\caption{Spectral function for different temperatures averaging over 10 configurations at the (a) VBM and (b) CBM. From the analysis of Fig.~\ref{fig:spectral_vs_Ncfgs}, the lineshift is well converged, while the shape of the spectral function as well as the width are not. (c) Combining the shifts from (a) and (b), we obtain the temperature dependence of the band gap in blue. The AHC on-shell analogous result is shown in red. (d) Linewidth as a function of temperature: as above, the peak width converges much more slowly with $N_\mathrm{cfgs}$.}
\label{fig:spectral_vs_T}
\end{figure}

\section{Results and discussion}
\label{sec:results}

As we mentioned in the Introduction, at the VBM, the width from AHC is 0, while in NPG there is a finite width. The difference in the spectral functions is shown in Fig.~\ref{fig:val_NP_vs_PT} (a). In AHC, the numerical non-zero width arises from the broadening parameter $\d$ and has no physical meaning. This can be verified in the inset of (b), where the width decreases linearly with $\d$ approximately towards a 0 intercept. 
With NPG instead, applying a linear fit, we obtain extrapolated intrinsic value of 97~meV for the VBM. 
The corresponding lineshifts extrapolate to $134$~meV at the VBM and
$-295$~meV at CB$_\Gamma$.
Since both versions of AHC are similar (Fig.~\ref{fig:val_NP_vs_PT}), it suggests a non-adiabatic non-perturbative calculation would also be similar to the adiabatic NPG.

Just as the lineshift is more accurate for on-shell AHC relative to Dyson-AHC because the energies in the denominator are consistent with each other (they are both bare), we also expect the linewidth to be more accurate for on-shell AHC.
From Dyson-AHC at CB$_\Gamma$, we obtain a width of $0.292$~eV, while the on-shell $\Sigma$ gives
$0.183$~eV, which is indeed closer to the NPG value of $0.123$~eV. A
difference of this size is nonetheless substantial, and part of it may
stem from the convergence of the AHC calculation, whose imaginary part is
the quantity most sensitive to the $\mbf{q}$-grid density and to the
choice of $\d$; a denser grid would be needed to establish how much of the
discrepancy is physical.

In order to determine the zero-point renormalization of the gap, we now look at the CBM. Diamond has an indirect gap, and the lowest conduction energy sits at around $\mbf{k}=3/4\,X$, where $X = 2\pi/a \,(1,0,0)$.
The picture there is analogous to the one at $\Gamma$, with the roles of the
conduction and valence bands reversed: NPG is again narrower than AHC and
decays to zero faster, with a less pronounced asymmetry (not shown). Note
that the valence band is doubly degenerate at this $\mbf{k}$-point while the
conduction band is not, so the peak height does not serve as a quick estimate
of the inverse width when comparing different states.

We also looked at the evolution of the spectral function with temperature at the VBM and CBM, as shown in Fig.~\ref{fig:spectral_vs_T} (a) and (b), respectively. In (c) we can see the temperature dependence of the band gap, and compare to the AHC self-energy at the bare energy. 
The zero-point renormalizations differ by $5\%$, and the values at 1500~K by
about $14\%$. The change between 0 and 1500~K is $330$~meV for NPG against
$423$~meV for AHC, a difference of $22\%$.
In (d), we also show the change of the width with temperature, although the numerical error may be significant since we only average over 10 configurations.

It is worth clarifying the precise relationship between the non-perturbative and perturbative treatments, since it is tempting to attribute their difference to higher-order irreducible diagrams. Expanding the exact NPG Green's function in powers of the atomic displacements and averaging reproduces, term by term, the standard perturbative series -- the Fan and DW self-energies at lowest order, higher-order diagrams beyond.
One might therefore expect that summing enough of these terms would recover
NPG. It does not. As discussed in Sec.~\ref{sec:LimsAHC}, such series are
generically asymptotic rather than convergent, so they cannot be summed to
the exact result at all. In particular, we showed earlier that adding more
diagrams cannot resolve the absence of broadening at the band extrema,
since this holds for every term of the series.
The essential distinction between NPG and PT is therefore not one of
diagram order but of analytic structure. This is why the self-consistent
Born approximation, which produces a branch cut rather than isolated
poles, does qualitatively better despite being built from the lowest-order self-energy.
For diamond, where the lineshift itself is already well
captured (on-shell) at lowest order, the main difference lies in the distribution of
spectral weight. In more strongly coupled materials, higher-order
contributions may become quantitatively important as well.

If the toy model and NPG differ so drastically, why do the perturbative and non-perturbative spectral functions at the $\Gamma$ conduction band (Fig.~\ref{fig:val_NP_vs_PT}) turn out to be overall similar rather than sharply contrasting? The resolution lies in the difference between an isolated level and an extended band. In the single-level toy model,
the truncated perturbative Green's function has only a discrete set of poles, and no width at all.
In an extended band, by contrast, the states span a continuous range of bare energies, so these individual poles are distributed over a finite interval and the resulting spectral function is spread out. A dense set of deltas across an interval approximates a smooth lineshape, which is why the perturbative and non-perturbative results agree far better here than the toy model would suggest. The agreement is nonetheless imperfect -- broadening a set of deltas is not the same as broadening a set of Gaussians, so perturbation theory recovers the overall envelope but misses the intrinsic per-state width, most visibly in the low-energy shoulder. More importantly, as we proved earlier, this rescue mechanism fails at band extrema, where the phase space for continuum broadening vanishes inside the gap: there are no bare states in the gap over which to distribute poles, so the toy-model pathology re-emerges while the non-perturbative treatment produces the correct spectral tail.

With these results in hand, we revisit the main limitations of the one-shot
approach of Ref.~\cite{Zacharias2020} discussed in the Introduction.
Fig.~\ref{fig:spectral_vs_Ncfgs}~(c) indicates that, for sufficiently large
SCs, comparable accuracy in the spectral shift can already be achieved with a
single arbitrary configuration. While our approach requires multiple
configurations, it offers important advantages. First, it yields the correct
spectral function, including the width, which the method of
Ref.~\cite{Zacharias2020} cannot capture. Second, Ref.~\cite{Zacharias2020}
estimates the gap by selecting, for the VBM, the highest-energy peak in the
distorted configuration (and conversely, the lowest for the CBM), which is
problematic for degenerate states. In a distorted configuration, degenerate
bands split, with each peak shifting to higher or lower energies (see
Fig.~\ref{fig:spectral_vs_Ncfgs}). Since by symmetry all members of the
multiplet have the same ensemble-averaged shift, systematically selecting the
extremal peak biases the estimate away from this common average, and no
individual band in a single distorted configuration serves as a reliable
reference for the gap.

Finally, we comment on a debated issue in the literature: in SC methods, does the self-energy of polar semiconductors diverge in the adiabatic approximation due to the $1/q$ dependence of the electron-phonon matrix elements? When omitting $\omega_{\mathbf{q}s}$ in the denominator and evaluating at $\omega=\varepsilon_{\mathbf{k}n}$, the self-energy in Eq.~\eqref{eq:Fan_ad} diverges.  This was noticed in PT in Ref.~\cite{Ponce2015}. Conversely, keeping the phonon frequencies in the denominators gives a well-behaved $\Sigma(\omega=\varepsilon_{\mathbf{k}n})$ in the weak-coupling limit.
Ultimately, in any approach based on the spectral function, since $\omega_\mathrm{QP}$ generally differs from the bare energy, the region where the spectral function has weight has no divergence (independently of how good the approximation is). While it is still possible that evaluating $\omega$ at the bare energy gives a divergence in the NPG self-energy for large SCs, we are typically not concerned with this region. From this point of view, there are no divergences in NPG in a polar material.\\

\section{Conclusions}
\label{sec:Conclusions}

In this work, we showed that the imaginary part of the perturbative self-energy at $T=0$
vanishes inside the bare gap at band extrema at any order, as a consequence of electron-number conservation at the vertices, which confines the intermediate states to the sectors with one electron more or one fewer, together with the absence of phonons at $T=0$.
Perturbation theory therefore produces no spectral broadening there,
regardless of how many diagrams are included. Producing this broadening
genuinely requires going beyond simple PT, for example through
self-consistent resummation or the non-perturbative Green's function approach
used here. For diamond, the lineshift itself is already well captured by
on-shell AHC; the main difference instead lies in the spectral broadening,
which stems not from higher-order diagrams but from the non-perturbative
nature of the method. Inserting a self-energy obtained from PT with bare electronic legs into the
Dyson equation should therefore be avoided, especially when the broadening, which cannot be reproduced at all at band extrema, is of interest.

The NPG spectral function is well converged with $N_s=8$, and with around 100
configurations -- significantly beyond what has been achieved in comparable SC
studies -- the linewidth converges to within $5\%$. This was made possible by GPU
acceleration, which reduced the computational cost by an order of magnitude.
The only quantity that requires extrapolation is the broadening parameter,
$\d\to0$. Furthermore, the lack of significant structural differences between the adiabatic and non-adiabatic
AHC spectral functions in diamond suggests that the adiabatic NPG framework
already captures the essential broadening physics.

NPG provides a method for determining the lineshift, linewidth, and
asymmetry of the spectral function within the adiabatic approximation. It also
serves as a useful diagnostic: when adiabatic NPG disagrees significantly with
adiabatic AHC, non-adiabatic PT is likely to be similarly unreliable. Future
applications of NPG to systems featuring strong electron-phonon coupling or
large anharmonicities may reveal non-perturbative effects that standard
perturbative expansions systematically miss.

\acknowledgments
J.P.N. is currently supported by the European Union under a Marie Sklodowska-Curie Individual Fellowship, Project GreenNP No. 101151380.
S.L. acknowledges funding from Horizon Europe MSCA Doctoral network grant n.101073486, EUSpecLab, funded by the European Union.
Computing time was provided by the Lucia Tier-1 HPC of the Fédération Wallonie-Bruxelles (Walloon Region grant agreement No. 1117545).

\appendix

\section{Numerical details}

Calculations were performed using ABINIT~\cite{verstraete25} and a carbon Troullier-Martins norm-conserving pseudopotential parameterized for the PBE functional, generated using the FHI98PP code (file downloaded from the abinit.org website in 2024). The potential was chosen because it is quite smooth, and accurate for diamond phonon properties. SC calculations with $N_s=8$ were performed at $\Gamma$, using a cutoff of 25 Ha, which converged the electronic spectral function within less than 1 meV, and 4096 bands (corresponding to 8 bands in the PC). Since the accessible $\mathbf{k}$-points are commensurate with the $N_s=8$ SC, and the CBM lies at approximately $3/4\,X$, which is commensurate with this SC, we computed the spectral function at this $\mathbf{k}$-point to determine the band-gap renormalization. This approximation is similar to that of Ref.~\cite{Ponce2025}, which looks at the lowest conduction energy state in a $9 \times 9 \times 9$ SC. For $N_s=2$ we used a $4 \times 4 \times 4$ Monkhorst-Pack grid, for $N_s= 4$ we used $2 \times 2 \times 2$, and for $N_s=6$ just $\Gamma$ as for $N_s=8$. To speed up calculations, especially regarding the $N_s=6$ and 8 SCs, we performed calculations on the Lucia cluster in Belgium, equipped with Nvidia GPUs. Ground state calculations for $N_s=8$ took about one hour using four nodes.

To calculate Eq.~\eqref{eq:Ginit}, in principle an infinite number of bands $n$ and $J$ are needed. However, we are only interested in energies around the top valence bands and lower conduction bands, which means energies $\vare^I_J$ on this range are mostly relevant. In AHC the sum over high-energy unoccupied bands converges notoriously
slowly. Although such bands barely affect the frequency dependence, which
is dominated by the poles in the energy range of interest, they contribute
a smooth background that displaces the whole curve.
Also, higher order bands $n$ might have a projection into states $|J \rangle$ in the energy range of interest. To guarantee sufficient bands are being included, we considered a $N_s=4$ SC, in order to be able to increase the number of bands (which became prohibitive for memory purposes with $N_s=8$), and look at the $\Gamma$ spectral function around the conduction band (noting that achieving convergence here necessitates a larger number of bands compared to the valence bands).
 
\begin{figure}[!t]
\includegraphics[scale=0.45]{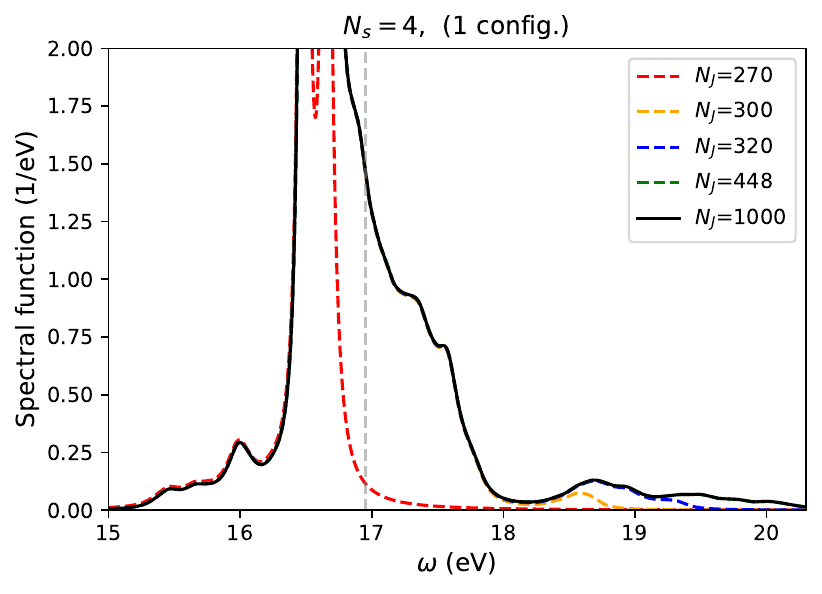} \\
(a) \\
\includegraphics[scale=0.45]{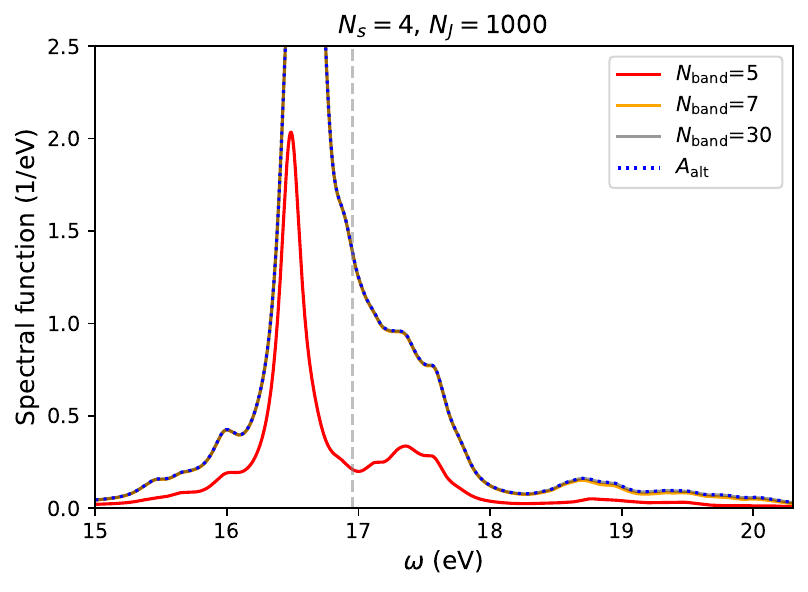} \\
(b)
\caption{Spectral function convergence, (a) for 1 configuration with respect to number of supercell bands  and (b) for fixed bands with respect to number of primitive cell bands. In (b), the dotted blue line $A_\mathrm{alt}$ is the spectral function of Eq.~\eqref{eq:A_allbands}, in which the sum over $n$ has been performed analytically. Going beyond 7 bands per PC (448 in the SC) does not change the spectral function in either convergence.}
\label{fig:conv_bands}
\end{figure}

In Fig.~\ref{fig:conv_bands}~(a), we fixed $N_\mathrm{band}=30$ and varied $N_J$, the number of $J$ bands. 
We can see that beyond $N_J=448$, which corresponds to 7 bands in the PC, the spectral function remains unchanged. Regarding the sum over $n$, the spectral function resulting from Eq.~\eqref{eq:Ginit} is 

\be
A(\o) = \sum_n \sum_J |\langle \mbf{k}n|J\rangle|^2 \d(\o - \vare_J),
\ee

\ni where we omit the $I$ index since here we are focusing on only one configuration. Using Eq.~\eqref{eq:overlap} and that $\sum_n |c_{n,\mbf{k}+\mbf{G}}|^2 = 1$ (since wavefunctions are normalized to 1), 

\vspace{-1cm}

\be
A(\o) = \sum_J |\tilde{c}_{J,\mbf{k}+\mbf{G}}|^2 \d(\o - \vare_J),
\label{eq:A_allbands}
\ee

\ni an expression for the spectral function where all bands have already been summed. In Fig.~\ref{fig:conv_bands}~(b), we fix $N_J=1000$, and see that $N_\mathrm{band}=7$ is virtually identical to the spectral function obtained with Eq.~\eqref{eq:A_allbands}. All calculations were done with $N_\mathrm{band}=10$ and $N_J=8 N_s^3$ bands (except for some buffer bands used by ABINIT to speed up convergence).
We are only using this expression now for convergence purposes, since this expression for the spectral function has the aforementioned issue of including a broadening (which inevitably has to be used to represent the delta function in the last equation) in the 0th order term.

\section{No spectral broadening inside the gap in PT}
\label{app:nogap}

\subsection{Hamiltonian and conventions}
\label{sec:conventions}

We consider $H = H_0 + V$, with

\be
H_0 = \sum_{\mbf{k}m} \vare_{\mbf{k}m}\, c^\dagger_{\mbf{k}m}
c_{\mbf{k}m} + \sum_{\mbf{q}s} \o_{\mbf{q}s}\,
\Big(a^\dagger_{\mbf{q}s} a_{\mbf{q}s} + \tfrac{1}{2}\Big),
\ee

\ni and the electron-phonon vertex $V$ built from the first-order matrix
elements $g^{\mbf{q}s}_{knm}$ of the main text,

\be
V = \frac{1}{\sqrt{N_{\mbf{q}}}}\sum_{\mbf{k}\mbf{q}}\sum_{m m' s}
g^{\mbf{q}s}_{m m'}(\mbf{k})\,
c^\dagger_{\mbf{k}+\mbf{q}m}\, c_{\mbf{k}m'}\,
\big(a_{\mbf{q}s} + a^\dagger_{-\mbf{q}s}\big).
\label{eq:vertex}
\ee

\ni The argument below
uses only two properties of the vertex: it conserves the electron number,
and it does not conserve the phonon number. The second-order vertex,
quadratic in the atomic displacements, shares both properties, so including
it changes nothing that follows. Diagrams containing closed electron loops
describe electron-mediated phonon renormalization and anharmonicity (a loop
with $L$ phonon legs contributes to the $L$th-order force constants); we
work with fixed phonons whose frequencies already include the electronic
response, such as those obtained with DFPT here, and therefore exclude such
diagrams to avoid double counting. Their inclusion would in any case not
alter the conclusions (see the remark at the end of Appendix~\ref{sec:zero_temperature}).

\subsection{Zero temperature}
\label{sec:zero_temperature}

At $T=0$ we determine where $\Im\mathrm{m}\,\Sigma$ can be nonzero within
time-ordered perturbation theory. This loses no generality: at any finite
order the time-ordered and retarded self-energies have identical real
parts, and imaginary parts equal up to an overall sign for $\o$ below the
chemical potential~\cite{FetterWalecka}, so one vanishes where the other
does.

The object to analyze is the time-ordered Green's function of the
interacting ground state $|\Psi_0\rangle$, $H|\Psi_0\rangle =
\mathcal{E}_0|\Psi_0\rangle$,
\be
i G_{\mbf{k}n}(t-t') = \langle \Psi_0 | T\, c_{\mbf{k}n}(t)\,
c^\dagger_{\mbf{k}n}(t') | \Psi_0 \rangle ,
\label{eq:Gdef}
\ee

\ni with the operators in the Heisenberg picture. The wavevector is a
spectator throughout and is suppressed from here on. Let $|\Phi_0\rangle$
be the unperturbed reference state, the filled valence determinant times
the phonon vacuum, with $H_0 |\Phi_0\rangle = E_0 |\Phi_0\rangle$, its
energy $E_0$ being distinct from the interacting $\mathcal{E}_0$ above. The
Gell-Mann--Low theorem~\cite{FetterWalecka} refers Eq.~\eqref{eq:Gdef} to
this state,

\be
i G_n(t-t') = \frac{\langle \Phi_0 | T\big[\, c_n(t)\, c^\dagger_n(t')\,
S \,\big] | \Phi_0 \rangle}{\langle \Phi_0 | S | \Phi_0 \rangle} ,
\label{eq:GML}
\ee

\ni with the $S$ matrix

\be
S = T \exp\Big[ -i \!\int\! d\tau\, e^{-\d|\tau|}\, V(\tau) \Big] ,
\ee

\ni the operators now evolving with $H_0$, $O(\tau) = e^{iH_0\tau} O
e^{-iH_0\tau}$, $V$ the electron-phonon potential of
Eq.~\eqref{eq:vertex}, and $\d \to 0^+$ the adiabatic switching rate. The
denominator of Eq.~\eqref{eq:GML} is a single number, independent of $t$
and $t'$; it therefore acquires no $\o$ dependence on Fourier
transformation, and plays no part in what follows.

We evaluate Eq.~\eqref{eq:GML} without contracting operators. Expanding
$S$ to order $r$, the integrand is symmetric in the interaction times, so
the $r!$ chronological orderings contribute equally and cancel the $1/r!$;
fixing $\tau_1 > \cdots > \tau_r$ and summing over the positions of $t$
and $t'$ within that chain, the operators are strictly ordered inside each
domain and $T$ may be dropped. Consider first an ordering with $t > t'$
and all the times $\tau_j$ between them. Collecting the exponentials and
writing $\mathcal{H}_0 = H_0 - E_0$, the matrix element is

\be
\langle \Phi_0 | c_n\, e^{-i\mathcal{H}_0 u_0}\, V\, e^{-i\mathcal{H}_0
u_1}\, V \cdots V\, e^{-i \mathcal{H}_0 u_r}\, c^\dagger_n | \Phi_0
\rangle ,
\label{eq:chain}
\ee

\ni with $u_0 = t - \tau_1$, $u_\alpha = \tau_\alpha - \tau_{\alpha+1}$ and
$u_r = \tau_r - t'$ the lengths of the intervals between successive
operators.

Which states can occupy those intervals follows from the first of the two
properties of $V$ noted in Appendix~\ref{sec:conventions}, conservation of
the electron number. Reading Eq.~\eqref{eq:chain} from right to left, that
is, from the earliest operator to the latest: $|\Phi_0\rangle$ carries
$N_e$ electrons; $c^\dagger_n$ raises the number to $N_e+1$; each $V$
leaves it unchanged; and $c_n$ restores $N_e$, so that the matrix element with $\langle \Phi_0 |$ need not vanish. Every state between the two external operators
therefore lies in the sector with one electron more than $|\Phi_0\rangle$,
and for the ordering $t < t'$ the same reading gives one electron fewer.
Any interaction times lying outside $[t', t]$ act before $c^\dagger_n$ or
after $c_n$, and the states between them remain in the $N_e$-electron
sector. The phonon occupations are subject to no such restriction, the
potential not conserving the phonon number.

Since $H_0$ is diagonal in the Fock basis, the eigenstates spanning the
two sectors reached by $c^\dagger_n$ and $c_n$ are the configurations

\be
|\chi\rangle = \Big(\prod_{i=1}^{p} c^\dagger_{c_i}\Big)
\Big(\prod_{j=1}^{h} c_{v_j}\Big) \Big(\prod_{\mbf{q}s}
\big(a^\dagger_{\mbf{q}s}\big)^{\ell_{\mbf{q}s}}\Big) |\Phi_0\rangle,
\label{eq:config}
\ee

\ni with  $p - h = \pm 1$, and $\ell_{\mbf{q}s} \in \{0, 1, 2, \dots\}$ the number of phonons in
mode $\mbf{q}s$: the filled valence bands with $p$ electrons added in
conduction states and $h$ removed from valence states, plus phonons. (The
identities $c^\dagger_{v}|\Phi_0\rangle = 0$ and $c_{c}|\Phi_0\rangle = 0$
express that no state exists with an electron added to a valence level or
removed from a conduction one.)

A resolution of the identity may now be inserted in each interval. For a
general ordering, let $k$ of the $r$ interaction times lie between $t'$
and $t$, with $u_0, \ldots, u_k$ the lengths of the intervals between the
external operators and $v_1, \ldots, v_{r-k}$ those of the intervals
outside them. Writing $\mathcal{H}_0 |\chi\rangle = \Omega_\chi
|\chi\rangle$ with $\Omega_\chi \equiv E_\chi - E_0$,
the chain becomes, in place of Eq.~\eqref{eq:chain},

\be
\sum_{\{\chi\}} \mathcal{N}_{\{\chi\}}
\prod_{\alpha=0}^{k} e^{-i\Omega_{\chi_\alpha} u_\alpha}
\prod_{\beta=1}^{r-k} e^{-i\Omega_{\chi'_\beta} v_\beta} ,
\label{eq:inserted}
\ee

\ni where $\chi_\alpha$ and $\chi'_\beta$ run over the configurations of
Eq.~\eqref{eq:config} and of the $N_e$-electron sector, respectively, and
$\mathcal{N}_{\{\chi\}}$ collects the matrix elements $\langle \chi | V |
\chi'' \rangle$ together with the two overlaps $\langle \Phi_0 | c_n |
\chi_0 \rangle$ and $\langle \chi_k | c^\dagger_n | \Phi_0 \rangle$. Since
no contraction is performed, neither propagators nor occupation factors
appear, and the intermediate states stand exposed. The matrix elements in
$\mathcal{N}_{\{\chi\}}$ are independent of $\o$ and can only delete terms
from the sum, never displace a pole.

The $\o$ dependence follows by Fourier transformation. Setting $t' = 0$,
the change of variables from $(t, \tau_1, \ldots, \tau_r)$ to the interval
lengths is triangular, so its Jacobian is unity, and the ordering
constraints become the independent conditions $u_\alpha > 0$, $v_\beta >
0$. Only the interior lengths sum to the external time, $t = \sum_\alpha
u_\alpha$, so $e^{i\o t}$ factorizes across those intervals alone and

\be
\begin{split}
\int_0^{\infty} \!\! dt\, e^{i \o t} & \int_{\mathcal{D}} \! d^r\tau
\prod_{\alpha=0}^{k} e^{-i \Omega_{\chi_\alpha} u_\alpha}
\prod_{\beta=1}^{r-k} e^{-i \Omega_{\chi'_\beta} v_\beta} \\
& = \prod_{\alpha=0}^{k} \frac{i}{\o - \Omega_{\chi_\alpha} + i\d}
\prod_{\beta=1}^{r-k} \frac{-i}{\Omega_{\chi'_\beta} - i\d} ,
\end{split}
\label{eq:denoms}
\ee

\ni with $\mathcal{D} = \{\, t > \tau_1 > \cdots > \tau_r > 0 \,\}$ the
ordered domain. Each interval between the two external operators thus
contributes one factor carrying $\o$, and each interval outside them a
constant, which cannot produce a pole. The opposite ordering $t < t'$
gives, by the same steps, factors $[\o + \Omega_\chi - i\d]^{-1}$.

The self-energy inherits this structure. It follows from
Eq.~\eqref{eq:denoms} by restricting to one-particle-irreducible
contributions and removing the two external legs $G_0(\o)$, so that the
denominators of $\Sigma$ are a subset of those of $G_n$. In the branch
surviving at zeroth order, $t > t'$ for $n$ unoccupied and $t < t'$ for
$n$ occupied, and for the orderings with every interaction time between
the external ones, the end overlaps are nonzero for a single basis state,
fixing $\Omega_{\chi_0} = \Omega_{\chi_k} = \vare_n$, so that the two
outermost factors are $[\o - \vare_n \pm i\d]^{-1} = G_0(\o)$ explicitly.

The excitation energy of any configuration of Eq.~\eqref{eq:config} is
Eq.~\eqref{eq:poleform} made concrete. For $p - h = 1$,

\be
\begin{split}
\Omega_\chi & = \sum_{i=1}^{p} \vare_{c_i} - \sum_{j=1}^{h} \vare_{v_j} +
\sum_{\mbf{q}s} \ell_{\mbf{q}s}\, \o_{\mbf{q}s} \\
& = \vare_{c_1} + \sum_{j=1}^{h}\big(\vare_{c_{j+1}} - \vare_{v_j}\big) +
\sum_{\mbf{q}s} \ell_{\mbf{q}s}\, \o_{\mbf{q}s} \\
& \geq\; \vare_{\mathrm{CBM}} + h\, E_{\mathrm{gap}},
\end{split}
\label{eq:bound}
\ee

\ni with all phonon terms non-negative, so that the corresponding denominators
vanish only at $\o \geq \vare_{\mathrm{CBM}}$. For $p - h = -1$ the same
pairing gives $\Omega_\chi \geq -\vare_{\mathrm{VBM}} + p\,
E_{\mathrm{gap}}$, and those denominators vanish at $\o =
-\Omega_\chi \leq \vare_{\mathrm{VBM}} - p\,
E_{\mathrm{gap}}$.

No denominator therefore vanishes for $\o$ strictly inside the gap, and
$\Im\mathrm{m}\,\Sigma(\o)$ is a sum of delta functions at these energies
(or derivatives thereof), identically zero throughout the gap at every
order. (The numerators are complex in general, but every sequence of intermediate
states is summed independently, so each chain occurs together with its
reverse; reversal conjugates $\mathcal{N}_{\{\chi\}}$, by hermiticity of
$V$, and leaves the denominators unchanged, since they are a product over
the same set of $\Omega_\chi$, so the sum is real wherever none of them
vanishes.) The bound holds for each
term of Eq.~\eqref{eq:inserted} individually, whether or not the
corresponding contribution is connected and one-particle irreducible;
those restrictions govern which terms enter $\Sigma$, not where their
poles lie. Equation~\eqref{eq:bound} further shows that configurations
with $h \geq 1$ (respectively $p \geq 1$), which produce the purely
electronic combinations of Eq.~\eqref{eq:poleform}, lie at least
$E_{\mathrm{gap}}$ beyond the band edges.

Comparison with Eq.~\eqref{eq:poleform} identifies which of the general
positions survive. Since $\ell_{\mbf{q}s} \geq 0$, at $T=0$ all phonon
frequencies enter a given denominator with the same sign, positive for the
states reached by adding an electron and negative for those reached by
removing one: the mixed signs permitted in general by the
non-conservation of the phonon number require a phonon to be absorbed,
but none is present.
Table~\ref{tab:weights} shows this explicitly for the crossed two-phonon-line diagram: the mixed-sign positions correspond to no state of the form Eq.~\eqref{eq:config}, and their weights vanish at $T=0$. The purely electronic position, having
$\ell_{\mbf{q}s} = 0$ throughout, attains the bound of
Eq.~\eqref{eq:bound} with $h=1$.

Nothing above singles out the diagrams with closed electron loops excluded
in Appendix~\ref{sec:conventions}, and their exclusion is accordingly a
matter of double counting rather than of validity: such a loop creates and
annihilates an electron-hole pair, so the states it generates are again of
the form Eq.~\eqref{eq:config}, with $p$ and $h$ each larger by the number
of pairs present. By Eq.~\eqref{eq:bound} every additional pair raises
$\Omega_\chi$ by at least $E_{\mathrm{gap}}$, placing these contributions
further from the gap rather than closer.

\subsection{Finite temperature and the weight of each pole}
\label{sec:finite_temperature}

At finite temperature the same positions acquire occupation weights.
Throughout, \emph{weight} means the occupation-dependent coefficient
multiplying a given pole position, and \emph{amplitude factor} the
remaining coefficient, built from the electron-phonon matrix elements and
the energy denominators. The structure of these weights rests on two facts.
First, from the Lehmann representation: inserting a complete set of exact
eigenstates of $H$ into each of the two time orderings, $c_n(t)\,
c^\dagger_n(t')$ and $c^\dagger_n(t')\,c_n(t)$, that make up the
time-ordered product, and relabeling the summation indices in the second,
both orderings contribute to the same delta function,

\be
A(\o) \propto \sum_{i,f} \big(p_i + p_f\big)\, \big|\langle f|\,
c^\dagger_{\mbf{k}n}\, |i\rangle\big|^2\, \delta\big(\o - (E_f -
E_i)\big),
\label{eq:lehmann}
\ee

\ni with $p_i$ the thermal weights and $|f\rangle$ containing one electron more
than $|i\rangle$; on the delta function, $E_f = E_i + \o$ gives the exact
relation $p_f = p_i\, e^{-\beta(\o - \mu)}$ (detailed balance), with $\mu$
anywhere in the gap.

Second, the weights follow from the intermediate-state expansion of
Appendix~\ref{sec:zero_temperature}. The object is now the thermal
average

\be
i G_n(t-t') = \frac{\mathrm{Tr}\big[ e^{-\beta H}\, T\{ c_n(t)\,
c^\dagger_n(t') \} \big]}{\mathrm{Tr}\, e^{-\beta H}} ,
\label{eq:Gthermal}
\ee

\ni which we evaluate by replacing $e^{-\beta H}$ with $e^{-\beta H_0}$ and
expanding the trace in the Fock basis. Instead of evaluating the
real-time operator chain in the single ground state $|\Phi_0\rangle$, it
is evaluated for each Fock state $|\Phi\rangle$ of $H_0$, of energy
$E_\Phi$, and then averaged over the noninteracting ensemble; the
Gell-Mann--Low construction of Eq.~\eqref{eq:GML}, specific to the
nondegenerate ground state, plays no role here. The insertion of complete
sets of $H_0$ eigenstates between successive operators proceeds exactly as
before, producing the same
real-frequency denominators with $|\Phi_0\rangle$ replaced by
$|\Phi\rangle$. 

The intermediate configurations are those of Eq.~\eqref{eq:config}, still
with $p - h = \pm 1$, but with the electron and phonon occupation changes
counted relative to $|\Phi\rangle$: adding an electron may now mean filling
a valence hole, and removing one may mean taking a conduction electron.
The operator identities
$c^\dagger_v|\Phi_0\rangle = 0$ and $c_c|\Phi_0\rangle = 0$ are replaced by
thermal averages $1 - f_{v}$ and $f_{c}$. Averaging over the noninteracting ensemble is an approximation: an exact
real-time treatment at finite temperature would replace $e^{-\beta H_0}$ by
$e^{-\beta H}$, adding vertices integrated over imaginary times in
$[0,\beta]$. These lie on a different branch of the contour from the real
times $t$ and $t'$, so the corresponding interval lengths do not appear in
$t-t'$; since $\o$ enters only through the Fourier transform of $t-t'$,
they yield $\o$-independent factors, modifying the weights at higher order
in $V$ but leaving the positions unchanged.
They are dropped in what follows; the
exact treatment is the Matsubara one described below.

Equation~\eqref{eq:bound} does not survive the replacement of
$|\Phi_0\rangle$ by $|\Phi\rangle$, and that is what admits weight inside
the gap. The bound rests on $|\Phi_0\rangle$
being the lowest eigenstate of $H_0$, with no phonons available to absorb,
no conduction electrons to remove and no valence holes to fill, so that
every occupation change raises $\Omega_\chi$, now measured from $E_\Phi$.
A thermally populated $|\Phi\rangle$ offers all three, contributions of
either sign enter $\Omega_\chi$, and positions forbidden at $T=0$, among
them the mixed
phonon signs of Table~\ref{tab:weights}, acquire nonzero weight.

Per changed mode, the squared matrix element contributes the eigenvalue of
$1 - c^\dagger_m c_m$ (an added electron) or $c^\dagger_m c_m$ (a removed
one), and of $a^\dagger_{\mbf{q}s} a_{\mbf{q}s} + 1$ (an emitted phonon) or
$a^\dagger_{\mbf{q}s} a_{\mbf{q}s}$ (an absorbed one); their thermal
averages are $1 - f_{m\mbf{k}}$, $f_{m\mbf{k}}$, $1 + n_{\mbf{q}s}$, and
$n_{\mbf{q}s}$; both the Fock space and $e^{-\beta H_0}$ factorize over
modes, so these averages do too. The position of a pole specifies
which modes change, so its weight is the product of one such factor per
changed mode; the $p_f$ term of Eq.~\eqref{eq:lehmann} contributes the same
product with the roles of initial and final state exchanged, i.e.\ with $f
\leftrightarrow 1-f$ and $n \leftrightarrow n+1$. The accompanying
amplitude factor does not depend on the occupations that enter the weight,
since the couplings and the energy denominators involve only the band
energies and phonon frequencies. For the Fan term, the position $\o =
\vare_{m\mbf{k}} + \o_{\mbf{q}s}$
is reached by adding an electron in $m\mbf{k}$ and emitting a phonon, of
weight $(1-f_{m\mbf{k}})(1+n_{\mbf{q}s})$, and by the reverse process, in
which an electron is removed from $m\mbf{k}$ and a phonon absorbed, of
weight $f_{m\mbf{k}} n_{\mbf{q}s}$; their sum is $1 + n_{\mbf{q}s} -
f_{m\mbf{k}}$. At $\o = \vare_{m\mbf{k}} - \o_{\mbf{q}s}$ emission and
absorption are exchanged and the sum is $n_{\mbf{q}s} + f_{m\mbf{k}}$,
recovering Eq.~\eqref{eq:Fan}; at $T=0$ all weights reduce to step
functions. At finite temperature the Fan
weight at a position inside the gap is $(1-f_{v\mbf{k}}) + n_{\mbf{q}s}$,
both terms exponentially small, which quantifies the suppression quoted in
the main text.

In the Matsubara formalism the same weights emerge less directly. There,
$\Sigma(\mbf{k}n, i\omega_j)$ at order $2L$ is built from the propagators

\be
\begin{split}
G_0(m, \mbf{k}', i\omega') & = \frac{1}{i\omega' - \vare_{\mbf{k}'m}}, \\
D_0(\mbf{q}s, i\nu) & = \frac{1}{i\nu - \o_{\mbf{q}s}} -
\frac{1}{i\nu + \o_{\mbf{q}s}},
\end{split}
\ee

\ni with $L$ internal bosonic frequencies $i\nu_1, \dots, i\nu_L$ summed over
and the external fermionic frequency continued at the end, $i\omega_j \to
\o + i\delta$. By frequency conservation, the frequency of an electron
segment is $i\omega_j$ shifted by the phonon frequencies that have come in
or gone out at the preceding vertices, $i\tilde\omega = i\omega_j - \sum_l
\sigma_l\, i\nu_l$ with $\sigma_l = \pm 1$ according to the direction; a
fermionic frequency shifted by bosonic ones remains fermionic, so
$e^{\beta i\tilde\omega} = -1$. (Consistency with Eq.~\eqref{eq:poleform}:
every electron denominator contains $i\omega_j$ once and with unit
coefficient, and no phonon denominator contains it, so
$\Sigma(i\omega_j; \{\vare + \Delta\vare\}) = \Sigma(i\omega_j -
\Delta\vare; \{\vare\})$; every pole
position must therefore shift rigidly by $\Delta\vare$, which forces the
signs in front of the electronic energies to sum to one.) Each frequency
sum evaluates to one residue per propagator pole, weighted by the Bose
function at the pole position:
$n_B(\o_{\mbf{q}s}) = n_{\mbf{q}s}$ at the positive phonon pole;
$n_B(-\o_{\mbf{q}s}) = -(1+n_{\mbf{q}s})$ at the negative one, whose
residue $-1$ makes the net factor $1 + n_{\mbf{q}s}$; and $n_B(\vare -
i\tilde\omega) = -n_F(\vare)$ at an electron pole. When real frequencies
have been substituted by earlier sums, composite factors such as
$n_F(\vare_{m\mbf{k}} \pm \o_{\mbf{q}s})$ appear, and a given final
denominator generally receives contributions from several distinct pole
sequences; the weights derived earlier in this subsection belong to the
\emph{collected} coefficient of each position. In crossed topologies a
further pole type arises: two electron propagators sharing a frequency
sum differ by another internal
frequency, and taking the pole of one inside the other leaves a denominator
$1/(i\nu' - \vare_a + \vare_b)$, free of the external frequency, whose own
pole contributes a Bose factor of an electronic energy difference. This is
the origin of the electronic differences in Eq.~\eqref{eq:poleform}; the
apparent divergence for $\vare_b \to \vare_a$ cancels between the two terms
in which the two propagators exchange roles.

As a complete worked example we consider the crossed diagram with two
phonon lines, whose frequency structure is

\be
\begin{split}
& G_0(1, i\omega_j{+}i\nu_1)\, G_0(2, i\omega_j{+}i\nu_1{+}i\nu_2)\,
G_0(3, i\omega_j{+}i\nu_2) \\
& \times\, D_0(1, i\nu_1)\, D_0(2, i\nu_2),
\end{split}
\ee

\ni with $\vare_1 \equiv \vare_{\mbf{k}+\mbf{q}_1 n_1}$, $\vare_2 \equiv
\vare_{\mbf{k}+\mbf{q}_1+\mbf{q}_2 n_2}$, $\vare_3 \equiv
\vare_{\mbf{k}+\mbf{q}_2 n_3}$, and $\o_a$, $n_a$ ($a = 1,2$) the two phonon
frequencies and occupations. Performing both sums and collecting by pole
position yields Table~\ref{tab:weights}: nine positions, each with the
weight predicted by the rule above. We have verified every entry
numerically to machine precision against the brute-force double Matsubara
sum of the raw product above, over independent parameter sets, including
the factorization of each collected weight, the amplitude factor being independent of the occupations appearing in the corresponding weight.
That such a decomposition
holds diagram by diagram at arbitrary order is the content of the
finite-temperature cutting rules~\cite{Weldon1983,Kobes1985,Kobes1986},
which we do not rely on, the claim of the main text being at $T=0$. Two
features of the table are worth noting: the mixed-sign two-phonon positions
vanish identically at $T=0$, since emitting one phonon while absorbing
another requires a thermal phonon; and the purely electronic position
carries no phonon occupations at all, both phonon propagators entering only
off shell, evaluated at electronic energy transfers, inside the amplitude
factor.

\begin{table}[t]
\begin{tabular}{lll}
\hline\hline
position & weight at finite $T$ & $T=0$ \\
\hline
$\vare_1 + \o_1$ & $(1{-}f_1)(1{+}n_1) + f_1 n_1$ & $1-f_1$ \\
$\vare_1 - \o_1$ & $(1{-}f_1)n_1 + f_1(1{+}n_1)$ & $f_1$ \\
$\vare_2 + \o_1 + \o_2$ & $(1{-}f_2)(1{+}n_1)(1{+}n_2) + f_2 n_1 n_2$ & $1-f_2$ \\
$\vare_2 - \o_1 - \o_2$ & $(1{-}f_2)n_1 n_2 + f_2(1{+}n_1)(1{+}n_2)$ & $f_2$ \\
$\vare_2 + \o_1 - \o_2$ & $(1{-}f_2)(1{+}n_1)n_2 + f_2 n_1(1{+}n_2)$ & $0$ \\
$\vare_2 - \o_1 + \o_2$ & $(1{-}f_2)n_1(1{+}n_2) + f_2(1{+}n_1)n_2$ & $0$ \\
$\vare_3 + \o_2$ & $(1{-}f_3)(1{+}n_2) + f_3 n_2$ & $1-f_3$ \\
$\vare_3 - \o_2$ & $(1{-}f_3)n_2 + f_3(1{+}n_2)$ & $f_3$ \\
$\vare_1 - \vare_2 + \vare_3$ & $(1{-}f_1)f_2(1{-}f_3) + f_1(1{-}f_2)f_3$ & $0$ or $1$ \\
\hline\hline
\end{tabular}
\caption{Pole positions of the crossed two-phonon-line diagram and their
collected weights, with $f_1, f_2, f_3$ and $n_1, n_2$ the occupations of
the three electronic states and two phonon modes; each weight multiplies an
amplitude factor independent of the occupations that enter it. In each
weight the first product is the direct process and the second its reverse,
related by $e^{-\beta(\o-\mu)}$. At $T=0$ the purely electronic position
survives only for the patterns (empty, occupied, empty) and (occupied,
empty, occupied) of the states $1,2,3$, which place it at least
$E_{\mathrm{gap}}$ beyond the band edges.}
\label{tab:weights}
\end{table}

\section{Degeneracy}
\label{sec:degeneracy}

We now analyze AHC and NPG in the degenerate and quasi-degenerate cases of a 2 band toy model (analogous to the one of the main section), in order to understand whether off-diagonal degenerate interactions can introduce significant width. For simplicity, here we only keep a linear coupling $g \,u$.

\subsection{Degenerate case}

In analogous fashion to the toy model of the main section, in this case we define

\be
H(u) = \begin{pmatrix} \vare_0 & g\, u \\ g^\ast u & \vare_0 \end{pmatrix}.
\ee

\ni As we determined earlier, the approximate expansion parameter is $||G_0 V(\s)|| = |\sqrt{\a}/x|$, just like in the intraband case (with no DW term).
The self-energy from AHC can be compactly written as $\Sigma = \langle V G_0 V \rangle$ (it can also be expressed in the usual AHC fashion), which gives a Green's function with diagonal components that coincide with the intraband case Eq.~\eqref{eq:G_intra_PT}.

On the other hand, the diagonal components of the distorted Green's function have the same form as Eq.~\eqref{eq:G_intra_PT}, but with $u$ instead of $\sigma$. This results in

\be
\begin{split}
A_{ii}(\o, u) & = -\frac{1}{\pi} \Im\mathrm{m} [G_{ii}(\o,u)] \\ 
& = \frac{1}{2} \Big[ \delta(\o - \vare_0 - gu) + \delta(\o - \vare_0 + gu) \Big],
\end{split}
\ee

\ni and finally

\be
\begin{split}
A_{ii}(\o) & = \int du \, P(u) A_{ii}(\o, u) \\ 
& = \frac{1}{|g|\sqrt{2\pi\sigma^2}} \exp\left(-\frac{(\o - \vare_0)^2}{2|g|^2\sigma^2}\right),
\end{split}
\ee

\ni which is identical to the intraband case Eq.~\eqref{eq:A_intra_NP}.

\subsection{Quasi-degenerate case}

Here we consider the same Hamiltonian as in the previous subsection, but with different diagonal energies $\vare_0 < \vare_1$,

\begin{equation}
H(u) = \begin{pmatrix} \vare_0 & g\,u \\ g^* u & \vare_1 \end{pmatrix}.
\end{equation}

\ni We now have 

\be
|| G_0 V(\s) || = \left| \frac{\alpha}{(\o' + d/2)(\o' - d/2)} \right|^{1/2} < 1,
\ee

\ni with $d = \vare_1 - \vare_0$ and $\o' = \o - (\vare_0+\vare_1)/2$. We now analyze separately the two regions: between the bands ($|\omega'| < d/2$) and outside them ($|\omega'| > d/2$). Outside the bands, convergence
requires $|\omega'| > \sqrt{(d/2)^2+\alpha}$. For weak coupling $\alpha \ll
d^2/4$, this is $|\omega'| > d/2 + \alpha/d$, placing $\omega$ at a distance
$\alpha/d$ beyond the nearest level. Between the bands, the condition becomes
$|\omega'| < \sqrt{(d/2)^2-\alpha} \approx d/2 - \alpha/d$, placing $\omega$
also away by $\alpha/d$ from the nearest level. Thus, for weak coupling,
$\omega$ must remain at least a distance $\alpha/d$ from either level,
regardless of the side. This corresponds to the non-degenerate case.

On the other hand, in the quasi-degenerate regime $\alpha \gtrsim d^2/4$
(strong coupling or small $d$), PT is uncontrolled between the bands, and
outside them $\omega$ must lie at a distance $\sqrt{\alpha}$ from the original
levels, which is linear in the coupling constant. This is a more stringent
condition than the non-degenerate case, where the distance $\alpha/d$ is
quadratic in the coupling constant.

Let us now analyze what happens with NPG.
Diagonalizing $H(u)$ at fixed $u$ yields the eigenvalues

\begin{equation}
\omega_{\pm}(u) = \frac{\vare_0 + \vare_1}{2} \pm \frac{1}{2}\sqrt{d^2 + 4 |g|^2 u^2}.
\end{equation}

\ni In the quasi-degenerate case $\a \gtrsim d^2/4$, both poles have to be considered for the variance, giving $d^2/4 + \a$. In the degeneracy limit $d \to 0$, we just recover the degenerate variance.  As $d$ gets larger, turning into the non-degenerate case, this expression does not represent a linewidth any longer since it corresponds to the distance between the poles, as opposed to their individual width. 

In the non-degenerate case, $\omega_-(u) \approx \vare_0 - \f{|g|^2 u^2}{d}$. Averaging over the Gaussian distribution $P(u)$, the pole position $\omega_-(u)$ fluctuates around its mean with variance $\langle \omega_-^2 \rangle - \langle \omega_- \rangle^2 = 2\alpha^2/d^2$, giving a spectral function of width

\begin{equation}
\Delta\omega \sim \frac{\sqrt{2}\,\alpha}{d}.
\end{equation}

\ni This has the standard second-order scaling.

\section{More details on PT, NPG and other methods}
\label{sec:PT_vs_NP_vs_SC}

We compare the series structure of PT, NPG, self-consistency, and the cumulant
expansion in the same solvable model, to show more clearly why PT does not lead
to the NPG result. Throughout this section we set $\alpha'=0$, i.e.\ we retain only the
linear coupling.

(i) \textit{AHC.} Inserting the lowest-order self-energy $\Sigma^\mathrm{Fan}=\alpha/x$, built from bare
propagators, into the Dyson series gives
\be
\begin{split}
G &= G_0 + G_0 \Sigma^\mathrm{Fan} G_0 + G_0 \Sigma^\mathrm{Fan} G_0 \Sigma^\mathrm{Fan} G_0 + \cdots \\
  &= \frac{1}{x} + \frac{\alpha}{x^3} + \frac{\alpha^2}{x^5} + \frac{\alpha^3}{x^7} + \cdots \\
  &= \frac{1}{x - \alpha/x} = \frac{x}{x^2 - \alpha}.
\end{split}
\label{eq:G_Fan}
\ee

\ni Since the self-energy is fixed, this is an ordinary geometric
series: each term is the previous one multiplied by the same factor
$\alpha/x^2$, and the series converges for $|\alpha/x^2|<1$. The result has only
two real poles, at $x=\pm\sqrt{\alpha}$, and zero linewidth at any order.

(ii) \textit{NPG.} Expanding $G(u)=1/(x-gu)$ and averaging term by term over
the Gaussian distribution of $u$ gives the series of
Eq.~\eqref{eq:G_avg_series},

\be
\begin{split}
G &= \frac{1}{x} + \frac{\alpha}{x^3} + \frac{3\alpha^2}{x^5} + \frac{15\alpha^3}{x^7} + \cdots \\
  &= \frac{1}{x} \sum_{k=0}^{\infty} (2k-1)!! \left(\frac{\alpha}{x^2}\right)^k,
\end{split}
\label{eq:G_NP}
\ee

\ni where the Gaussian moments $(2k-1)!! = 1,1,3,15,105,\ldots$ count the
number of ways to pair up $2k$ points. Comparing with Eq.~\eqref{eq:G_Fan}
identifies precisely what AHC keeps: of the $(2k-1)!!$ pairings at order $k$,
the Dyson chain retains a single one, and its geometric convergence is a
property of that subset, not of the full expansion, whose coefficients grow
factorially and yield zero radius of convergence.

(iii) \textit{Self-consistency.} Setting $\alpha'=0$ in the self-consistent
solution of the main text gives

\be
G = \frac{x-\sqrt{x^2-4\alpha}}{2\alpha},
\label{eq:G_SC}
\ee

\ni corresponding to the implicit equation $\Sigma =
\alpha/(x-\Sigma)$. Unlike Eq.~\eqref{eq:G_Fan}, this is not a series of
repeated insertions of a fixed self-energy: $\Sigma^\mathrm{SC}$ appears on
both sides, so each iteration nests the entire previous result inside a new
reciprocal, generating a continued fraction rather than a geometric series.
Writing $\mathcal{G}\equiv xG^\mathrm{SC}$ and $z\equiv\alpha/x^2$, the implicit
equation reduces to $\mathcal{G}=1+z\mathcal{G}^2$, precisely the defining functional equation
of the Catalan generating function. Its series expansion is therefore fixed
exactly,

\be
G^\mathrm{SC} = \frac{1}{x} + \frac{\alpha}{x^3} + \frac{2\alpha^2}{x^5} +
\frac{5\alpha^3}{x^7} + \frac{14\alpha^4}{x^9}+\cdots,
\ee

\ni with coefficients $C_k=\frac{1}{k+1}\binom{2k}{k}$, the Catalan
numbers~\cite{Catalan1838}
: the number of ways to pair up $2k$ points \textit{without crossings}. 
Self-consistency therefore
generates every rainbow diagram to infinite order, but no crossed one,
the exact count of all pairings being $(2k-1)!!$ as in
Eq.~\eqref{eq:G_avg_series}. Since $C_{k+1}/C_k \to 4$, this series
converges only for $|x| > 2\sqrt{\alpha}$, exactly up to the branch point
of the closed form, Eq.~\eqref{eq:G_SC}. Where it converges it is real, and
reproduces the perturbative result with no width at all: the broadening
lives entirely in the region $|x| < 2\sqrt{\alpha}$ that the series cannot
reach, and arises from the branch cut of the self-consistent solution
rather than from summing perturbative terms.

(iv) \textit{Cumulant.} A fourth route exponentiates the Fan self-energy in the
time domain, $G(t) = -i\theta(t)e^{-i\vare_0 t}e^{C(t)}$, with

\be
C(t) = \frac{1}{\pi}\int d\o'\,
       \frac{|\Im\mathrm{m}\Sigma^{\mathrm{Fan}}(\vare_0+\o')|}{\o'^2}
       \left(e^{-i\o' t} + i\o' t - 1\right).
\ee

\ni \ni Here $\Im\mathrm{m}\Sigma^{\mathrm{Fan}} = -\pi\alpha\,\delta(\o')$, and since the
kernel vanishes as $\o'^2$ at the origin it cancels the $1/\o'^2$, leaving a
finite integrand from which $\delta(\o')$ picks out

\be
C(t) = -\frac{\alpha}{2}t^2.
\ee

\ni Fourier transforming, $A(\o) = -\tfrac{1}{\pi}\Im\mathrm{m} G$ reduces to the even
cosine integral

\be
\begin{split}
A(\o) & = \frac{1}{2\pi}\int_{-\infty}^{\infty} dt\,
        e^{i(\o-\vare_0)t}\,e^{-\alpha t^2/2} \\
      & = \frac{1}{\sqrt{2\pi\alpha}}\,
        e^{-(\o-\vare_0)^2/2\alpha},
\end{split}
\ee

\ni recovering the exact NPG result, Eq.~\eqref{eq:A_intra_NP}, as expected
from the exactness of the cumulant under the conditions of
Ref.~\cite{Dunn1975}, which the present model obeys trivially.

\bibliography{bibliography}

\end{document}